\documentclass[a4paper,10pt]{article}
\pdfoutput=1
\usepackage[utf8]{inputenc}
\usepackage{cite}
\usepackage{graphicx}
\usepackage{amsmath}
\usepackage{gensymb}

\def\lsim{\raise0.3ex\hbox{$\;<$\kern-0.75em\raise-1.1ex\hbox{$\sim\;$}}}
\def\gsim{\raise0.3ex\hbox{$\;>$\kern-0.75em\raise-1.1ex\hbox{$\sim\;$}}}
\title{Decays of a NMSSM CP-odd Higgs in the low-mass region}
\author{Florian Domingo}
\date{{\em Instituto de F\'isica Te\'orica (UAM/CSIC), Universidad Aut\'onoma de Madrid, Cantoblanco, E-28049 Madrid, Spain}\\[.4em]
{\em Instituto de F\'isica de Cantabria (CSIC-UC), E-39005 Santander, Spain}}

\begin{document}

\maketitle
\vspace{-6cm}\rightline{IFT-UAM/CSIC-16-142}\vspace{6cm}
\begin{abstract}
A popular regime in the NMSSM parameter space involves a light CP-odd Higgs $A_1$. This scenario has consequences for e.g.\ light singlino Dark Matter annihilating
in the $A_1$-funnel. In order to confront the pseudoscalar to experimental limits such as flavour observables, Upsilon decays or Beam-Dump experiments, it is
necessary to control the interactions of this particle with hadronic matter and derive the corresponding decays. The partonic description cannot be relied upon
for masses close to $m_{A_1}\sim1$~GeV and we employ a chiral lagrangian, then extended to a spectator model for somewhat larger masses, to describe the interplay
of the CP-odd Higgs with hadrons. Interestingly, a mixing can develop between $A_1$ and neutral pseudoscalar mesons, leading to substantial hadronic decays and a
coupling of $A_1$ to the chiral anomaly. Additionally, quartic $A_1$-meson couplings induce tri-meson decays of the Higgs pseudoscalar. We investigate these effects
and propose an estimate of the Higgs widths for masses below $m_{A_1}\lsim3$~GeV. While we focus on the case of the NMSSM, our results are applicable to a large class of models.
\end{abstract}

\section{The NMSSM and a light CP-odd Higgs}

While the hunt for physics beyond the Standard Model at the high-energy frontier continues at the LHC or in Dark Matter experiments -- with disappointing results 
so far --, new physics may still have a few surprises in store in the low-mass region. Axion-phenomenology is a classical example of such effects in the limit of
light less-than-weakly-coupled particles. In the following, we consider another case of comparatively light state occurring in the context of the Next-to-Minimal 
Supersymmetric Standard Model (NMSSM) \cite{Ellwanger:2009dp}, a well-motivated extension of the Standard Model (SM).

In the NMSSM, the CP-odd Higgs sector (ignoring the Goldstone boson) consists of two degrees of freedom -- a doublet component $A^0$, 
comparable to the MSSM pseudoscalar, and a singlet state $A^0_S$. Both mix at tree-level according to the following mass-matrix:
\begin{align}\label{massmat}
 {\cal M}_{\mbox{\tiny CP-odd}}^2&=\begin{pmatrix}
                             \frac{2\lambda s}{\sin{2\beta}}(A_{\lambda}+\kappa s)&\lambda v(A_{\lambda}-2\kappa s)\\
                             \lambda v(A_{\lambda}-2\kappa s)&-3\kappa s A_{\kappa}+\frac{\lambda v^2\sin{2\beta}}{2s}(A_{\lambda}+4\kappa s)
                            \end{pmatrix}\\ &=P\cdot\mbox{diag}(m^2_{A_1},m^2_{A_2})\cdot P^T\hspace{2cm}P\equiv\begin{pmatrix}\cos\theta_P&-\sin\theta_P\\ \sin\theta_P&\cos\theta_P\end{pmatrix}\nonumber
\end{align}
where $\lambda$ and $\kappa$ are parameters from the superpotential, $A_{\lambda}$ and $A_{\kappa}$, parameters from the soft supersymme\-try-breaking lagrangian, $v=(2\sqrt{2}G_F)^{-1/2}$, $\tan\beta$ and $s$, 
doublet and singlet vacuum expectation values or related quantities. Here, we have been considering the $Z_3$-conserving NMSSM explicitly. However, $Z_3$-violating terms as well
as radiative corrections can be incorporated in this picture with limited effort. The states $A_i=P_{i1}\,A^0+P_{i2}\,A^0_S$ are ordered in mass: $m_{A_1}<m_{A_2}$.
A light state is regarded as natural -- i.e.\ as a pseudo-Nambu-Goldstone boson -- in two specific limits of the NMSSM parameter space:
\begin{itemize}
 \item For $\kappa\ll\lambda$, the Higgs potential is approximately invariant under a $U(1)$ Peccei-Quinn symmetry. In this case, the doublet component of 
 $A_1$ -- through which this particle couples to SM matter -- is given by $P_{11}=-\frac{v}{2s}\sin{2\beta}/\sqrt{1+\frac{v^2}{4s^2}\sin^2{2\beta}}$.
 \item For $A_{\lambda},\ A_{\kappa}\to0$, another approximate $U(1)$ symmetry appears, which can be related to the R-symmetry. In this case, $P_{11}=\frac{v}{s}\sin{2\beta}/\sqrt{1+\frac{v^2}{s^2}\sin^2{2\beta}}$.
\end{itemize}
However, the light pseudoscalar may also result from an `accidental' arrangement of the NMSSM parameters.

While a MSSM pseudoscalar might still be comparatively light \cite{Bechtle:2016kui} (see also \cite{Domingo:2015eea} for a discussion in the context of the NMSSM),
the hypothesis of a dominantly doublet light CP-odd Higgs -- already constrained (at least indirectly) by LEP \cite{Barate:2003sz} -- is under increasing pressure from 
LHC searches \cite{Aad:2014kga,Khachatryan:2015qxa} or flavour transitions, due to the correlation of doublet masses -- implying that there exist a light CP-even and a 
comparatively light pair of charged Higgs states as well, all phenomenologically more conspicuous at colliders. 
Still, the situation is different for a singlet or a mixed pseudoscalar $A_1$, since the mentioned correlation dissipates and the light CP-odd Higgs becomes 
largely independent from the rest of the Higgs sector. From the perspective of LEP (or $e^+e^-$-colliders in general), the direct production of a CP-odd Higgs proves
difficult as the tree-level couplings to electroweak gauge-bosons vanish. The possibility of a light CP-odd NMSSM state -- with mass $\lsim10$~GeV -- thus 
appeared in the pre-LHC era as a phenomenologically appealing and realistic scenario: see e.g.\ \cite{Dobrescu:2000jt,Dobrescu:2000yn,Dermisek:2006wr,Morrissey:2008gm}. 
Nevertheless, limits from low-energy observables, e.g.\ flavour transitions \cite{Hiller:2004ii,Domingo:2007dx,Heng:2008rc,Andreas:2010ms,Domingo:2015wyn} or 
bottomonium decays and spectroscopy \cite{Drees:1989du,SanchisLozano:2002pm,SanchisLozano:2003ha,SanchisLozano:2005di,McElrath:2005bp,SanchisLozano:2006gx,
Dermisek:2006py,Fullana:2007uq,Hodgkinson:2008ei,McKeen:2008gd,Domingo:2008rr,Domingo:2009tb,Dermisek:2010mg,Domingo:2010am}, apply in this low-mass region and constrain, in particular, the 
coupling of $A_1$ to down-type fermions. With the start of the LHC, several direct or indirect production modes of the light CP-odd Higgs have been considered 
\cite{Almarashi:2012ri,Rathsman:2012dp,Cerdeno:2013cz,Bomark:2014gya,Bomark:2015fga,Conte:2016zjp}. Yet, the Higgs discovery at the LHC \cite{Aad:2012tfa,Chatrchyan:2012xdj} 
considerably reduces the scope of the phenomenology associated to a light $A_1$: when kinematically allowed, the Higgs-to-Higgs decay $H[125]\to2A_1$ -- with $H[125]$ denoting the 
observed state at $\sim125$~GeV -- could naively dominate the standard decay channels, which would have implied suppressed rates of $H[125]$ in the Run-I (and Run-II).
As the observed Higgs characteristics demonstrate the success of the standard search channels, the $H[125]\to2A_1$ decay width must therefore be small. This can be 
realized -- in certain 
limits or due to accidental cancellations -- and, in this extent, a light NMSSM $A_1$ may coexist with a CP-even state at $\sim125$~GeV that retains roughly SM-like characteristics -- 
hence a suitable candidate for $H[125]$. Such a scenario thus remains phenomenologically viable. Nevertheless, the condition of a suppressed $H[125]\to2A_1$ induces constraints on the NMSSM
parameter space, which 
have been discussed in e.g. \cite{Cao:2013gba,Domingo:2015eea}. Additionally, ATLAS and CMS have searched explicitly for $H[125]\to2A_1$ with final states including leptons \cite{Aad:2015oqa,Khachatryan:2015wka,Khachatryan:2015nba}.

Beyond its consequences for the Higgs sector, the hypothetical existence of a light NMSSM pseudoscalar may lead to other phenomenologically interesting effects. 
In the context of singlino Dark Matter \cite{Cerdeno:2004xw,Belanger:2005kh,Cerdeno:2007sn,Hugonie:2007vd}, the light-$A_1$ funnel may ensure a sufficiently large 
annihilation cross-section, yielding the correct relic density \cite{Vasquez:2010ru,Cao:2011re,Vasquez:2012hn,Han:2014nba,Ellwanger:2014dfa,Han:2015zba,Cerdeno:2015jca}. In this sense, the light CP-odd Higgs
scenario retains a clear motivation. Another application \cite{Ellwanger:2016wfe} would address the $17$~MeV excess in $^8Be$ transitions \cite{Krasznahorkay:2015iga}.

Despite the interest that the light CP-odd Higgs scenario has raised in the literature, a relative shadow continues to veil our knowledge of the decays of this particle in the very
low mass range $m_{A_1}\lsim2m_{\tau}$. There, the partonic description, summarized in e.g.\ \cite{Dermisek:2010mg}, predicts largely dominant strongly-interacting
final states, such as $gg$ or $s\bar{s}$. This picture has been sensibly criticized by \cite{Dolan:2014ska}: close to the confinement scale, the partonic approach
is no longer reliable and \cite{Dolan:2014ska} recommends an effective description of the hadronic decays based on the perturbative spectator model. Among the consequences
of the latter choice, $m_{A_1}=3\,m_{\pi^0}\sim0.4$~GeV becomes the lower limit where hadronic final states are relevant. Moreover, hadronic channels then seem largely superseeded by the
$A_1\to\mu^+\mu^-$ width. Still, this description in \cite{Dolan:2014ska} misses at
least one effect that can substantially affect the decays: the CP-odd Higgs shares its quantum numbers with (some of) the mesons, which induces a mixing among these 
states. In other words, the light CP-odd Higgs acquires a mesonic component -- via its interaction with quarks -- and the latter may well dominate the decays of this 
particle. In this sense, the impact of hadronic physics
extends below the tri-pion threshold, at least down to $m_{A_1}\sim m_{\pi^0}$. This mixing effect has already been noted in the context of heavy quarkonia 
\cite{Drees:1989du,Domingo:2009tb} and its impact on $A_1$-decays at the $b\bar{b}$-threshold was highlighted in \cite{Domingo:2011rn}. In the very low-mass range,
\cite{Ellwanger:2016qax} suggested that the hadronic decays of $A_1$ may resemble those of the meson that is closest in mass, while \cite{Domingo:2016unq} estimated the 
mixing with the mesons in the formalism of Partially-Conserved Axial Currents (PCAC). As a consequence of this confused situation for the pseudoscalar decays, the 
phenomenology of this particle at low-mass remains largely speculative and the interplay of constraints cannot be consistently applied.

In this paper, we aim at shedding some light into this question and propose an estimate of the NMSSM pseudoscalar decays in the $m_{A_1}\lsim3$~GeV range. Due to the intrinsic difficulty of a quantitative
description of hadronic phenomena and the corresponding large uncertainties, this evaluation has no ambition beyond that of providing an educated guess for the $A_1$
decay widths and branching fractions and, while the derived picture may seem more reliable than the partonic approach, we should not dismiss the possibility of 
sizable deviations. In the following section, we shall summarize the formalism describing the interactions of a light CP-odd Higgs with the mesons, relevant at
masses below $\lsim1$~GeV. Then, we will derive the $A_1$ decays in this mass range. Finally, we will attempt to extrapolate the hadronic decays of the pseudoscalar 
up to the $c\bar{c}$ threshold using the perturbative spectator approach, before coming to a short conclusion.

As a final word before starting with the actual description of the pseudoscalar interactions at low mass, we stress that our results apply beyond the NMSSM, in any singlet or 
doublet extension of the SM containing a light pseudoscalar Higgs state: all that is necessary in order to extend our discussion to such cases amounts to replacing the explicit NMSSM couplings to quarks,
photons and gluons by their analogues in the corresponding model.

\section{From the partonic lagrangian to the mesonic interactions}

The purpose of this section consists in summarizing the formalism leading to the inclusion of a light pseudoscalar in the non-linear Sigma model for the mesons.

\subsection{Partonic lagrangian below the \boldmath $c\bar{c}$ threshold}

We consider a NMSSM CP-odd Higgs with mass below the $c\bar{c}$ threshold. The other relevant fields at low-energy include the up, down and strange quarks, the muon and 
electron, as well as the photon and gluons. The interactions of the pseudoscalar with these fields may be summarized in the following effective lagrangian:
\begin{align}\label{LA1}
 {\cal L}_{A_1}=&\frac{\imath\, P_{11}}{\sqrt{2}v}A_1\left\{m_u\tan^{-1}\beta\,\bar{u}\gamma_5u+m_d\tan\beta\,\bar{d}\gamma_5d+m_s\tan\beta\,\bar{s}\gamma_5s+m_{\mu}\tan\beta\,\bar{\mu}\gamma_5\mu+m_e\tan\beta\,\bar{e}\gamma_5e\right\}\nonumber\\
 &+\frac{\alpha}{4\pi}C_{\gamma}\,A_1 F_{\mu\nu}\tilde{F}^{\mu\nu}+\frac{\alpha_s}{4\pi}C_g\,A_1 G_{\mu\nu}^a\tilde{G}^{a\,\mu\nu}
\end{align}
Here, we have confined to the operators of lowest-dimension for the $A_1$-interactions with each type of field: dimension $4$ for the fermions and dimension $5$ for the
gauge bosons. We have kept the tree-level expression of the fermionic couplings, though part of the radiative corrections may be incorporated within $P_{11}$ -- defined by Eq.(\ref{massmat}) and corresponding 
to the proportion of doublet-component in $A_1$. $F_{\mu\nu}$ and $G_{\mu\nu}^a$ denote the field-strength tensors for the photonic and gluonic fields respectively;
$\tilde{F}_{\mu\nu}$ and $\tilde{G}_{\mu\nu}^a$ are their dual.

The couplings $C_{\gamma}$ and $C_g$ are generated by loops of heavy fermions ($t$, $b$, $\tau$, $c$
and charginos); heavy scalars and gauge bosons are known not to contribute, due to non-renormalization theorems. Since we regard the light quarks and leptons as 
`active' fields, we do not include their radiative contribution in $C_{\gamma}$ and $C_g$: for the leptons, this effect could be added straightforwardly; in the 
case of light quark contributions, however, inclusion at the partonic level should be reputed unreliable. At the one-loop level:
\begin{align}
 C_{\gamma}=&-\frac{P_{11}}{2\sqrt{2}\,v}\left\{\frac{N_cQ_u^2}{\tan\beta}\left[{\cal F}\left(\frac{m_t^2}{m^2_{A_1}}\right)+{\cal F}\left(\frac{m_c^2}{m^2_{A_1}}\right)\right]
 +N_cQ_d^2\tan\beta\,{\cal F}\left(\frac{m_b^2}{m^2_{A_1}}\right)+Q_e^2\tan\beta\,{\cal F}\left(\frac{m_{\tau}^2}{m^2_{A_1}}\right)\right\}\nonumber\\
 &-\frac{Q_{\chi}^2}{2\sqrt{2}}\sum_{i=1}^2\frac{1}{m_{\chi_i^{\pm}}}\left[\lambda P_{12}U_{i2}V_{i2}-g\,P_{11}\left(\cos\beta\,U_{i1}V_{i2}+\sin\beta\,U_{i2}V_{i1}\right)\right]{\cal F}\left(\frac{m_{\chi_i^{\pm}}^2}{m^2_{A_1}}\right)\\
 C_g=&-\frac{P_{11}}{4\sqrt{2}\,v}\left\{\frac{1}{\tan\beta}\left[{\cal F}\left(\frac{m_t^2}{m^2_{A_1}}\right)+{\cal F}\left(\frac{m_c^2}{m^2_{A_1}}\right)\right]
 +\tan\beta\,{\cal F}\left(\frac{m_b^2}{m^2_{A_1}}\right)\right\}\nonumber\\
 N_c=&3\hspace{2mm};\hspace{2mm}Q_u=\frac{2}{3}\hspace{2mm};\hspace{2mm}Q_d=-\frac{1}{3}\hspace{2mm};\hspace{2mm}Q_e=Q_{\chi}=-1\hspace{2mm};\hspace{2mm}{\cal F}(x)=2x\log^2\left[\frac{\sqrt{1-4x}-1}{\sqrt{1-4x}+1}\right]\nonumber
\end{align}
Beyond the obvious notations for the SM fermions, we have introduced the chargino masses $m_{\chi_i^{\pm}}$ and diagonalizing matrices $U$, $V$: we refer the reader
to appendix A of \cite{Ellwanger:2009dp} for the details of the conventions. We remind that the logarithm in the definition of ${\cal F}$ is taken in its complex sense.

Finally, we define the axial currents associated to the light quarks:
\begin{equation}
 J_A^{a\,\mu}=(\bar{q})^T\gamma^{\mu}\gamma_5\lambda^a(q)\hspace{1cm};\hspace{1cm}(q)\equiv(u,d,s)^T
\end{equation}
where $\lambda^a$ are the Gell-Mann matrices acting in flavour space -- in particular $\lambda^3\equiv\mbox{diag}(1,-1,0)/\sqrt{2}$, $\lambda^8\equiv\mbox{diag}(1,1,-2)/\sqrt{6}$ and 
$\lambda^9\equiv\mbox{diag}(1,1,1)/\sqrt{3}$ -- and normalized to $\mbox{Tr}[\lambda^a\lambda^b]=\delta^{ab}$. We note that the $A_1$ 
couplings to the light quarks may be related to the divergences of the neutral currents \cite{Domingo:2016unq}:
\begin{multline}
{\cal L}_{A_1}\ni\frac{P_{11}}{4v}A_1\left\{(\tan^{-1}\beta-\tan\beta)\partial_{\mu}J^{3\,\mu}_A+\frac{1}{\sqrt{3}}(\tan^{-1}\beta-\tan\beta)\partial_{\mu}J^{8\,\mu}_A\right.\\\left.+\sqrt{\frac{2}{3}}(\tan^{-1}\beta+2\tan\beta)\partial_{\mu}J^{9\,\mu}_A\right\}+\ldots
\end{multline}
We also remind the coupling of the photon and gluon to the chiral anomalies:
\begin{align}
 \partial_{\mu}J_A^{a\,\mu}&=\imath(\bar{q})^T\gamma_5\{\lambda^a,m_q\}(q)+\frac{\alpha_s}{4\pi}\mbox{Tr}[\lambda^a]G_{\mu\nu}^a\tilde{G}^{a\,\mu\nu}+\frac{2\alpha N_c}{4\pi}\mbox{Tr}[\lambda^aQ_q^2]F_{\mu\nu}\tilde{F}^{\mu\nu}\nonumber\\
 &\partial_{\mu}J_A^{3\,\mu}=\sqrt{2}\imath\left[m_u\bar{u}\gamma_5u-m_d\bar{d}\gamma_5d\right]+\frac{\alpha}{4\pi}\sqrt{2}F_{\mu\nu}\tilde{F}^{\mu\nu}\\
 &\partial_{\mu}J_A^{8\,\mu}=\sqrt{\frac{2}{3}}\imath\left[m_u\bar{u}\gamma_5u+m_d\bar{d}\gamma_5d-2m_s\bar{d}\gamma_5s\right]+\frac{\alpha}{4\pi}\sqrt{\frac{2}{3}}F_{\mu\nu}\tilde{F}^{\mu\nu}\nonumber\\
 &\partial_{\mu}J_A^{9\,\mu}=\frac{2}{\sqrt{3}}\imath\left[m_u\bar{u}\gamma_5u+m_d\bar{d}\gamma_5d+m_s\bar{d}\gamma_5s\right]+\frac{\alpha_s}{4\pi}\sqrt{3}G_{\mu\nu}^a\tilde{G}^{a\,\mu\nu}+\frac{\alpha}{4\pi}\frac{4}{\sqrt{3}}F_{\mu\nu}\tilde{F}^{\mu\nu}\nonumber
\end{align}
with $m_q=\mbox{diag}(m_u,m_d,m_s)$ and $Q_q=\mbox{diag}(\frac{2}{3},-\frac{1}{3},-\frac{1}{3})$ the quark mass and charge matrices.

\subsection{Chiral lagrangian}
The dynamics of the mesons is well described -- at lowest order in a momentum expansion -- by a non-linear sigma model known as the Chiral Perturbation Theory 
($\chi$PT) \cite{Rosenzweig:1979ay,DiVecchia:1980yfw,Kawarabayashi:1980dp,Kawarabayashi:1980uh,Gasser:1984gg}. This formalism relies on a controlled breaking of the axial symmetries and proves remarkably 
predictive. Though refinements including higher-dimension terms are possible \cite{Gasser:1984gg}, they lead to a fast increase of the number of free low-energy parameters and we
shall confine to the simplest approach below. A recurrent endeavour of the 1980's consisted in estimating the couplings of a 
hypothetically light SM Higgs boson -- or a 2HDM CP-even state -- to the hadronic sector \cite{Ellis:1975ap,Shifman:1979eb,Vainshtein:1980ea,Voloshin:1985tc,
Voloshin:1986hp,Ruskov:1987jg,Grinstein:1988yu,Chivukula:1988gp,Chivukula:1989ds,Leutwyler:1989tn,Leutwyler:1989xj,Dawson:1989kr,Pich:1991dg}. Some attention was also paid to the
case of a pseudoscalar \cite{Grzadkowski:1992av,Chang:2008np}, and has persisted till today at least from the perspective of axion physics (see e.g.\ 
\cite{diCortona:2015ldu} for a recent reference). In the following, we aim at summarizing the key ingredients that intervene in the description of the
interactions of a light CP-odd Higgs with the meson sector.

The starting point of $\chi$PT rests with the observation that the QCD lagrangian for the light quarks $(q)$ preserves the axial symmetry -- characterized by the transformation $(q)\mapsto U[\alpha_a](q)$, with 
$U[\alpha_a]\equiv\exp[\imath \alpha_a\lambda^a\gamma_5]$ -- up to the mass term $M_q$ and the electromagnetic interaction. In our case, the Yukawa couplings to the light CP-odd Higgs -- see Eq.(\ref{LA1}) -- can be 
incorporated within the mass matrix:
\begin{multline}
 {\cal L}_{q}=(\bar{q})^T\left\{\imath\gamma^{\mu}D_{\mu}-M_q[A_1]\right\}(q)\hspace{0.5cm};\hspace{0.5cm}D_{\mu}(q)\equiv\left(\partial_{\mu}-\imath g_s T^a G^a_{\mu}-\imath e Q_q A_{\mu}\right)(q)\\
 M_q[A_1]\equiv\mbox{diag}\left[m_u\left(1-\frac{\imath P_{11}}{\sqrt{2}v\tan\beta}A_1\right),m_d\left(1-\frac{\imath P_{11}\,\tan\beta}{\sqrt{2}v}A_1\right),m_s\left(1-\frac{\imath P_{11}\,\tan\beta}{\sqrt{2}v}A_1\right)\right]
\end{multline}
with $g_s$ the strong coupling constant, $T^a$ the Gell-Mann matrices in colour-space, $e$ the elementary electric charge, $Q_q$ the quark-charge matrix (as defined above), $G^a_{\mu}$ the gluon field and $A_{\mu}$ the photon field.

Since the strong interaction triggers the formation of quark condensates at low energy, we shift our attention from the fundamental $3$ to the $\bar{3}\times3$ representations of $U(3)_{\mbox{\tiny flavour}}$, the mesonic octet 
and singlet $\Sigma_{ij}\sim\left<\bar{q}_i\gamma_5q_j\right>$. Then, the lowest-order effective lagrangian for $\Sigma$ preserving the axial (and vectorial) symmetry up to $M_q$ and $Q_q$ reads:
\begin{equation}\label{chiPT}
 {\cal L}_{\chi}=\mbox{Tr}\left\{D_{\mu}\Sigma\,D^{\mu}\Sigma^{\dagger}+\frac{B}{2}\left[M_q[A_1]\Sigma+\Sigma^{\dagger}M_q[A_1]^{\dagger}\right]\right\}+\frac{C}{2}\left(\partial_{\mu}K^{\mu}\right)^2+\imath\partial_{\mu}K^{\mu}\left[\frac{1}{2}\mbox{Tr}\log\Sigma-\imath C_gA_1\right]
\end{equation}
Here, $D_{\mu}\Sigma\equiv\partial_{\mu}\Sigma-\imath eA_{\mu}[Q_q,\Sigma]$, $B$ and $C$ are coupling constants and $\partial_{\mu}K^{\mu}\sim\frac{\alpha_s}{4\pi}G_{\mu\nu}^a\tilde{G}^{a\,\mu\nu}$ is an auxilliary field designed 
to mimic the gluon coupling to the $U(1)_A$ anomaly \cite{Rosenzweig:1979ay}. In other words, considering the axial currents ${\cal J}_{\mu}^a\equiv\imath\mbox{Tr}\left\{\partial_{\mu}\Sigma\left\{\lambda^a,\Sigma^{\dagger}\right\}-\left\{\lambda^a,\Sigma\right\}\partial_{\mu}
\Sigma^{\dagger}\right\}$, one obtains:
\begin{equation}
 \partial_{\mu}{\cal J}^{9\,\mu}\ni\sqrt{3}\partial_{\mu}K^{\mu}\sim\frac{\sqrt{3}\alpha_s}{4\pi}G_{\mu\nu}^a\tilde{G}^{a\,\mu\nu}
\end{equation}
Similarly, the gluonic coupling of $A_1$ is accounted for in Eq.(\ref{chiPT}) by the $\partial_{\mu}K^{\mu}$ term. The minimization condition for $\partial_{\mu}K^{\mu}$ provides the chiral lagrangian:
\begin{equation}
 \tilde{\cal L}_{\chi}=\mbox{Tr}\left\{D_{\mu}\Sigma\,D^{\mu}\Sigma^{\dagger}+\frac{B}{2}\left[M_q[A_1]\Sigma+\Sigma^{\dagger}M_q[A_1]^{\dagger}\right]\right\}-\frac{1}{2C}\left[C_gA_1+\frac{\imath}{2}\mbox{Tr}\log\Sigma\right]^2
\end{equation}

We finally introduce the pion fields $\pi_a$ as $\Sigma\equiv\frac{f_{\pi}}{2}\exp\left[\frac{\imath\sqrt{2}}{f_{\pi}}\pi_a\lambda^a\right]$ and expand the lagrangian in terms of these:
\begin{multline}\label{lowen}
 \tilde{\cal L}_{\chi}\simeq\frac{1}{2}\left\{D_{\mu}\pi_aD^{\mu}\pi_a+\frac{B\,P_{11}}{v}A_1\pi_a\mbox{Tr}\left[\lambda^a\tilde{M}_q\right]-\frac{B}{f_{\pi}}\pi_a\pi_b\mbox{Tr}\left[\lambda^a\lambda^bm_q\right]\right.\\
 \left.-\frac{BP_{11}}{3vf_{\pi}^2}A_1\pi_a\pi_b\pi_c\mbox{Tr}\left[\lambda^a\lambda^b\lambda^c\tilde{M}_q\right]+\frac{B}{6f_{\pi}^3}\pi_a\pi_b\pi_c\pi_d\mbox{Tr}[m_q\lambda^a\lambda^b\lambda^c\lambda^d]
 -\frac{1}{C}\left[C_gA_1-\frac{1}{f_{\pi}}\sqrt{\frac{3}{2}}\pi_9\right]^2\right\}+\ldots
\end{multline}
where $m_q\equiv M_q[0]$ and $\tilde{M}_q\equiv\mbox{diag}\left[\frac{m_u}{\tan\beta},m_d\tan\beta,m_s\tan\beta\right]$. We observe that this procedure generates mass terms for the mesons, a mass shift for $A_1$, mixing terms
between the mesons and $A_1$ as well as quartic interaction terms involving $A_1$ and three pions -- the conservation of CP excludes a cubic coupling. Our derivation of the decays of the light pseudoscalar will be based on this
simple lagrangian. Using the pion equations of motion, it is possible to check that, as in the quark model, the couplings of $A_1$ to the hadronic sector follow the PCAC, i.e.
\begin{equation}
 \tilde{\cal L}_{\chi}\ni\frac{P_{11}A_1}{4v}\left\{\left(\tan^{-1}\beta-\tan\beta\right)\partial^{\mu}{\cal J}^3_{\mu}+\frac{1}{\sqrt{3}}\left(\tan^{-1}\beta-\tan\beta\right)\partial^{\mu}{\cal J}^8_{\mu}
 +\sqrt{\frac{2}{3}}\left(\tan^{-1}\beta+2\tan\beta\right)\partial^{\mu}{\cal J}^9_{\mu}\right\}
\end{equation}
This could have been chosen as an equivalent Ansatz for the $A_1$ interactions. Yet, the previous formalism has allowed us to include the $A_1$ coupling to gluons in the low-energy picture as well.

So far we have omitted Wess-Zumino-Witten terms describing the pion coupling to photons \cite{Wess:1971yu,Witten:1983tw} -- and restoring the corresponding contribution to the anomaly:
\begin{equation}
 {\cal L}_{WZW}\simeq-\frac{\sqrt{2}N_c\alpha}{4\pi f_{\pi}}\pi_a\mbox{Tr}\left[Q_q^2\lambda^a\right]F_{\mu\nu}\tilde{F}^{\mu\nu}+\ldots
\end{equation}
We now add this piece to Eq.(\ref{lowen}).

\subsection{Low-energy coupling constants -- meson masses}
The low-energy lagrangian of Eq.(\ref{lowen}) has left us with five combinations of couplings to determine at low-energy: $\frac{Bm_u}{f_{\pi}}$, $\frac{Bm_d}{f_{\pi}}$, $\frac{Bm_s}{f_{\pi}}$, $f_{\pi}$, $C$.
$f_{\pi}\simeq93$~MeV is the pion decay constant: it determines the pion coupling to the anomaly, hence its diphoton decay, from which it is extracted. The other parameters are usually obtained from the mass matrix of the mesons:
from Eq.(\ref{lowen}), $\left[{\cal M}^2_{\pi}\right]^{ab}=\frac{B}{f_{\pi}}\mbox{Tr}\left[\lambda^a\lambda^bm_q\right]+\frac{3}{2f_{\pi}^2C}\delta^{a9}\delta^{b9}$. We may thus identify:
\begin{itemize}
 \item three pions with diagonal mass $m^2_{\pi}\equiv\frac{B}{2f_{\pi}}(m_u+m_d)\simeq(135~\mbox{MeV})^2$; the electromagnetic interaction generates an additional mass contribution to the charged pions 
 $\pi^{\pm}=\frac{\pi_1\mp\imath\pi_2}{\sqrt{2}}$; in principle, the neutral pion $\pi_3$ mixes with other neutral states ($\pi_8$, $\pi_9$), but the mixing terms $\propto m_u-m_d$ are numerically small so that we may neglect them
 at this level.
 Thus $\frac{Bm_u}{f_{\pi}}\simeq\frac{Bm_d}{f_{\pi}}\simeq m^2_{\pi}$.
 \item a pair of charged kaons $K^{\pm}=\frac{\pi_4\mp\imath\pi_5}{\sqrt{2}}$ with mass $m_{K^{\pm}}^2=\frac{B}{2f_{\pi}}(m_u+m_s)\simeq(494~\mbox{MeV})^2$ and a pair of neutral kaons $K^0,\bar{K}^0=\frac{\pi_6\mp\imath\pi_7}{\sqrt{2}}$ with mass 
 $m_{K^{0}}^2=\frac{B}{2f_{\pi}}(m_d+m_s)\simeq(498~\mbox{MeV})^2$. Therefore, $\frac{Bm_s}{f_{\pi}}\simeq m_{K^{0}}^2+m_{K^{\pm}}^2-m^2_{\pi}$.
 \item The neutral $\pi_8$ and $\pi_9$ mix according to the following matrix:
 \begin{multline}
  \begin{bmatrix}
   m_{\pi_8}^2&\Delta\\ \Delta&m^2_{\pi_9}
  \end{bmatrix}
\hspace{0.5cm};\hspace{0.5cm}m_{\pi_8}^2\equiv\frac{B}{6f_{\pi}}(m_u+m_d+4m_s)\hspace{0.5cm};\hspace{0.5cm}m_{\pi_9}^2\equiv\frac{B}{3f_{\pi}}(m_u+m_d+m_s)+\frac{3}{2f_{\pi}^2C}\hspace{0.5cm};\hspace{0.0cm}\\\Delta\equiv\frac{B}{3\sqrt{2}f_{\pi}}(m_u+m_d-2m_s)
 \end{multline}
The corresponding mass states are the $\eta=\cos\theta_{\eta}\,\pi_8-\sin\theta_{\eta}\,\pi_9$ and $\eta'=\sin\theta_{\eta}\,\pi_8+\cos\theta_{\eta}\,\pi_9$, with masses of $m_{\eta}\simeq548$~MeV and $m_{\eta'}\simeq958$~MeV 
and a mixing angle $\theta_{\eta}$ of order $-13\degree$ (see e.g.\ \cite{Ambrosino:2009sc,Ricciardi:2012xu,Pham:2015ina}). The values of $m_{\pi_8}^2\simeq(575~\mbox{MeV})^2$
and $\Delta\simeq-(370~\mbox{MeV})^2$ are in rough agreement with what we could expect in view of the pion and kaon masses. $m^2_{\pi_9}\simeq(940~\mbox{MeV})^2$ provides $\frac{3}{2Cf_{\pi}^2}\simeq m^2_{\pi_9}-\frac{1}{3}(m_{K^{0}}^2+m_{K^{\pm}}^2+m^2_{\pi})\simeq m^2_{\pi_9}-\frac{1}{2}(m_{\pi_8}^2+m_{\pi}^2)\simeq(850~\mbox{MeV})^2$
\end{itemize}
This fully determines the low-energy parameters that we employ in the following.

\section{Decays of a CP-odd Higgs from the chiral lagrangian}
The chiral lagrangian of Eq.(\ref{lowen}) that we have derived in the previous section characterizes the interactions of the Higgs pseudoscalar with the mesons. At low masses
$m_{A_1}\lsim1$~GeV, it should prove a reliable guide for the hadronic decays of the CP-odd Higgs. Additionally, the leptonic and photonic interactions can be taken directly
from Eq.(\ref{LA1}). In this section, we derive the associated phenomenology in this low-mass regime. In order to smoothen the transition with the higher-mass region, we shall display the 
Higgs decays up to $m_{A_1}\sim1.5$~GeV in numerical applications.

\subsection{\boldmath $A_1$-meson mixing}
Eq.(\ref{lowen}) shows that the interactions of the pseudoscalar with the hadronic sector generate a mixing of $A_1$ with $\pi_3$, $\pi_8$ and $\pi_9$:
\begin{equation}
 \tilde{\cal L}_{\chi}\ni-\frac{1}{2}(\pi_3,\pi_8,\pi_9,A_1){\cal M}^2_{\mbox{\small mix}}\begin{pmatrix}\pi_3\\\pi_8\\\pi_9\\A_1\end{pmatrix}\hspace{0.5cm};\hspace{0.5cm}
 {\cal M}^2_{\mbox{\small mix}}=\begin{bmatrix}
                       m^2_{\pi} & \frac{1}{\sqrt{3}}\delta & \sqrt{\frac{2}{3}}\delta & \delta m^2_3\\
                       \frac{1}{\sqrt{3}}\delta & m^2_{\pi_8} & \Delta & \delta m^2_8\\
                       \sqrt{\frac{2}{3}}\delta & \Delta & m^2_{\pi_9} & \delta m^2_9\\
                       \delta m^2_3 & \delta m^2_8 & \delta m^2_9 & \bar{m}^2_{A_1}
                      \end{bmatrix}
                      \end{equation}\begin{displaymath}
     \begin{cases}\delta\equiv\frac{B}{2f_{\pi}}(m_u-m_d)\simeq0\\
       \delta m^2_3\equiv-\frac{BP_{11}}{2\sqrt{2}v}\left(\frac{m_u}{\tan\beta}-m_d\tan\beta\right)=-\frac{f_{\pi}P_{11}}{2\sqrt{2}v}\left[m^2_{\pi}\left(\tan^{-1}\beta-\tan\beta\right)+\delta\left(\tan^{-1}\beta+\tan\beta\right)\right]\\
       \delta m^2_8\equiv-\frac{BP_{11}}{2\sqrt{6}v}\left(\frac{m_u}{\tan\beta}+m_d\tan\beta-2m_s\tan\beta\right)\\ \hfill=-\frac{f_{\pi}P_{11}}{2\sqrt{2}v}\left[-\sqrt{3}\tan\beta m^2_{\pi_8}+\frac{m^2_{\pi}}{\sqrt{3}}\left(\tan^{-1}\beta+2\tan\beta\right)+\frac{\delta}{\sqrt{3}}\left(\tan^{-1}\beta-\tan\beta\right)\right]\\
       \delta m^2_9\equiv-\frac{BP_{11}}{2\sqrt{3}v}\left(\frac{m_u}{\tan\beta}+m_d\tan\beta+m_s\tan\beta\right)-\sqrt{\frac{3}{2}}\frac{C_g}{f_{\pi}C}\\ \hfill=-\frac{f_{\pi}P_{11}}{2\sqrt{2}v}\left[\sqrt{\frac{3}{2}}\tan\beta m^2_{\pi_8}+\frac{m^2_{\pi}}{\sqrt{6}}\left(2\tan^{-1}\beta+\tan\beta\right)+\sqrt{\frac{2}{3}}\delta\left(\tan^{-1}\beta-\tan\beta\right)\right]\\ \hfill-\sqrt{\frac{2}{3}}f_{\pi}C_g\left[m^2_{\pi_9}-\frac{1}{3}(m^2_{\pi_8}+m^2_{\pi})\right]\\
       \bar{m}^2_{A_1}=m^2_{A_1}+\frac{1}{C}C_g^2
      \end{cases}\nonumber
\end{displaymath}
We note that the mass-shift of $A_1$ resulting from the gluon coupling -- $\frac{1}{C}C_g^2$ -- is typically tiny: sub-MeV$^2$.

The diagonalization of ${\cal M}^2_{\mbox{\small mix}}=O\cdot \mbox{diag}[\tilde{m}^2_{\pi},\tilde{m}^2_{\eta},\tilde{m}^2_{\eta'},\tilde{m}^2_{A_1}]\cdot O^T$ -- where $O$ is the orthogonal transition matrix -- defines the 
mass-states $(\tilde{\pi}^0,\tilde{\eta},\tilde{\eta}',\tilde{A_1})^T=O\cdot(\pi_3,\pi_8,\pi_9,A_1)^T$. If the presence of the Higgs pseudoscalar were to affect
drastically the characteristics of the pions, then this mixing scenario would be phenomenologically difficult to defend. In practice, however, the mixing is naturally small.
We note indeed that the off-diagonal elements of ${\cal M}^2_{\mbox{\small mix}}$ are suppressed by a prefactor $\frac{f_{\pi}}{v}\sim5\cdot10^{-4}$. Moreover, they explicitly 
involve the doublet component of $A_1$ -- $P_{11}$ controls the interactions of the pseudoscalar with SM-matter --, which tends to be suppressed, i.e.\ $P_{11}^2\ll1$ 
in the realistic cases: the light $A_1$ is likely to intervene as a dominantly singlet state. Consequently, the $A_1$-meson mixing has a negligible impact on the mesons. 
In \cite{Domingo:2016unq}, for instance, it was shown that the pseudoscalar could approach the pion mass within MeV wihout affecting the pion decays as long as the $A_1$-$\pi_3$ 
mixing remained below $\sim4\%$. Similarly, the mass-shift associated with the mixing remains tiny. Another type of limit proceeds from the conversion of hadrons to
$A_1$ via mixing. For instance, $K^+\to\pi^+\pi^0$ induces $K^+\to\pi^+A_1$. In the case of a sizable $A_1\to\mbox{invisible}$ decay, the $K^+\to\pi^+\nu\bar{\nu}$
searches would place a limit on the $A_1-\pi_3$ mixing. Similar bounds proceed from leptonic decays (or leptonic transitions mediated by an off-shell $A_1$). The
interference with loop-generated flavour-changing $A_1$ couplings would have to be considered, however. Additionally, experimental cuts require some attention as e.g.\ the 
vicinity of the pion mass is occasionally discarded in order to avoid the pion background. We shall not enter into a detailed discussion here as the $A_1$ decays need first be derived. 
We may thus safely assume $\tilde{\pi}^0\sim\pi_3$, $\tilde{\eta}\sim\eta$ 
and $\tilde{\eta}'\sim\eta'$, although a narrower check would be in order when $m_{A_1}$ is very near a meson mass. We shall discuss this further when we compute
the leptonic decay widths.
On the other hand, the impact of the mixing on the phenomenology of $A_1$ is sizable because the couplings of this field to SM particles are suppressed in the 
same proportion as the mixing: in other words, even a small meson component in $\tilde{A}_1$ could dominate the genuine $A_1$ amplitudes. In the following, we 
thus focus on the corresponding state\footnote{We employ $3$, $8$, $9$, $A$ indices to refer to the $\pi_3$, $\pi_8$, $\pi_9$, $A_1$ components; similarly, indices $\eta$ and $\eta'$
correspond to a rotation of angle $\theta_{\eta}$ with respect to the $\pi_8$ and $\pi_9$ components.}:
\begin{equation}
 \tilde{A}_1=O_{A3}\,\pi_3+O_{A8}\,\pi_8+O_{A9}\,\pi_9+O_{AA}\,A_1=O_{A3}\,\pi_3+O_{A\eta}\,\eta+O_{A\eta'}\,\eta'+O_{AA}\,A_1
\end{equation}
In the limit where the mixing angles remain small -- which is almost systematically fulfiled and ensures that the mesons do not receive a sizable Higgs component --, 
these matrix elements can be approximated by:
\begin{align}
 &O_{AA}\simeq1&
 &O_{A3}\simeq\frac{\delta m^2_3}{m_{A_1}^2-m^2_{\pi}}\\
 &O_{A\eta}\simeq\frac{\cos\theta_{\eta}\,\delta m^2_8-\sin\theta_{\eta}\,\delta m^2_9}{m_{A_1}^2-m^2_{\eta}}&
 &O_{A\eta'}\simeq\frac{\sin\theta_{\eta}\,\delta m^2_8+\cos\theta_{\eta}\,\delta m^2_9}{m_{A_1}^2-m^2_{\eta'}}\nonumber 
\end{align}
Nevertheless, in numerical applications, we will retain the full numerical evaluation of these objects.

\begin{figure}[thb!]
\begin{center}
\includegraphics[width=0.7\textwidth]{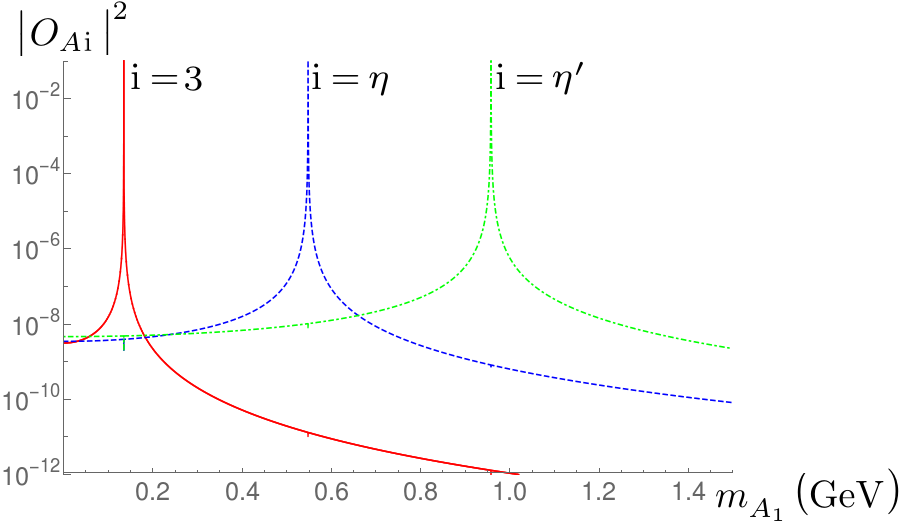}
\caption{$A_1$-meson mixing for $P_{11}=0.03$, $\tan\beta=10$. The full red line corresponds to $|O_{A3}|^2$, the dashed blue line, to $|O_{A\eta}|^2$ and the dot-dashed
green line, to $|O_{A\eta'}|^2$, i.e.\ the $A_1$-pion, $A_1-\eta$ and $A_1-\eta'$ mixing respectively.
\label{fig:mixing}}
\end{center}
\end{figure}
We plot these mixing angles in Fig.\ref{fig:mixing} for $P_{11}=0.03$, $\tan\beta=10$. The mixings prove very small (below $\sim10^{-8}$) on the whole mass-range, except when
$m_{A_1}$ is in the immediate vicinity of a meson mass. The impact of the mixing on the mesonic state will thus remain negligible.

Then, any decay amplitude of the pseudoscalar may be decomposed as:
\begin{equation}
{\cal A}[\tilde{A}_1\to X]=O_{AA}\,{\cal A}[A_1\to X]+O_{A3}\,{\cal A}[\pi_3\to X]+O_{A\eta}\,{\cal A}[\eta\to X]+O_{A\eta'}\,{\cal A}[\eta'\to X]
\end{equation}
where all the amplitudes on the right-hand side should be worked out for the $\tilde{A}_1$ kinematics, though.

\subsection{Photonic decay}
The diphoton decay is one of the channels where the mixing with the mesons has the most dramatic effects for the CP-odd Higgs, due to the large, anomaly-driven diphoton decays 
of the mesons. The amplitudes can be worked out from Eqs.(\ref{LA1}) and (\ref{lowen}) -- for photons with momenta $p_1$ and $p_2$ and polarizations $\varepsilon(p_1)$ and $\varepsilon(p_2)$:
\begin{align}
 &{\cal A}[A_1\to \gamma\gamma]=\frac{\alpha}{4\pi}\left(C_{\gamma}+\delta C_{\gamma}^{e,\mu}\right)\varepsilon^{\mu\nu\rho\sigma}\left[p_{1\mu}\varepsilon_{\nu}(p_1)-p_{1\nu}\varepsilon_{\mu}(p_1)\right]\left[p_{2\rho}\varepsilon_{\sigma}(p_2)-p_{2\sigma}\varepsilon_{\rho}(p_2)\right]\nonumber\\
 &\hspace{1cm}\delta C_{\gamma}^{e,\mu}=-\frac{P_{11}}{2\sqrt{2}v}Q_e^2\tan\beta\left\{{\cal F}\left(\frac{m_{e}^2}{m^2_{A_1}}\right)+{\cal F}\left(\frac{m_{\mu}^2}{m^2_{A_1}}\right)\right\}\\
 &{\cal A}[\Pi\to \gamma\gamma]=\frac{\alpha}{4\pi}C_{\gamma}[\Pi]\,\varepsilon^{\mu\nu\rho\sigma}\left[p_{1\mu}\varepsilon_{\nu}(p_1)-p_{1\nu}\varepsilon_{\mu}(p_1)\right]\left[p_{2\rho}\varepsilon_{\sigma}(p_2)-p_{2\sigma}\varepsilon_{\rho}(p_2)\right]\hspace{0.2cm};\hspace{0.2cm}\Pi=\pi_3,\eta,\eta'\nonumber\\
 &\hspace{1cm}C_{\gamma}[\pi_3]=\frac{\sqrt{2}N_c}{f_{\pi}}\mbox{Tr}[\lambda^3Q_q^2]\hspace{0.5cm};\hspace{0.5cm}C_{\gamma}[\eta]=\frac{\sqrt{2}N_c}{f_{\pi}}\left(\cos\theta_{\eta}\,\mbox{Tr}[\lambda^8Q_q^2]-\sin\theta_{\eta}\,\mbox{Tr}[\lambda^9Q_q^2]\right)\hspace{0.5cm};\hspace{0.2cm}\nonumber\\
 &\hspace{1cm}C_{\gamma}[\eta']=\frac{\sqrt{2}N_c}{f_{\pi}}\left(\sin\theta_{\eta}\,\mbox{Tr}[\lambda^8Q_q^2]+\cos\theta_{\eta}\,\mbox{Tr}[\lambda^9Q_q^2]\right)\nonumber
 \end{align}
\begin{figure}[t]
\begin{center}
\includegraphics[width=0.7\textwidth]{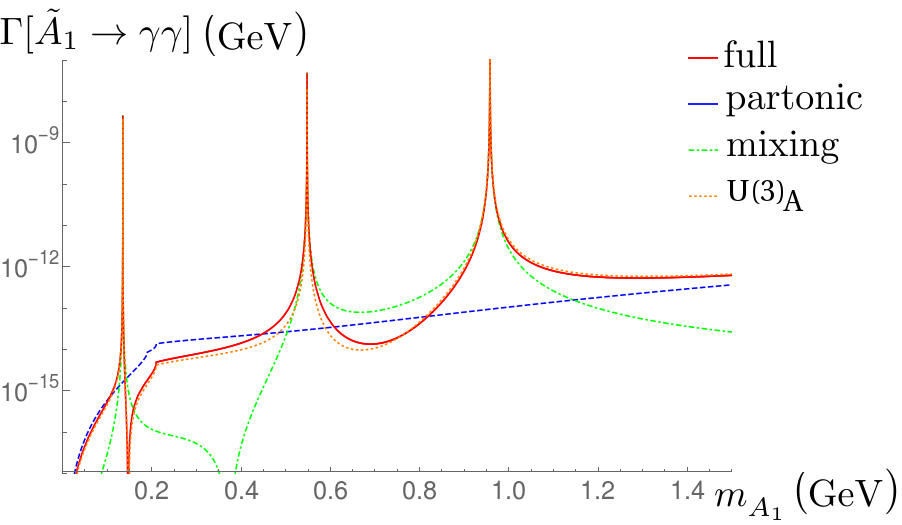}
\caption{Diphoton width of the mostly-$A_1$ state. $P_{11}=0.03$, $\tan\beta=10$, $M_2=\mu_{\mbox{\tiny eff}}=500$~GeV, $\lambda=0.3$. In (dashed) blue is the partonic width (including a partonic
strange quark with mass $0.095$~GeV). The (dotdashed) green curve corresponds to the width mediated by the meson mixing (i.e.\ neglecting $C_{\gamma}$). The full result of
Eq.(\ref{diphotonwidth}), employing phenomenological estimates of the pion-photon couplings, is in red (full line). For the orange dotted curve, we have imposed strict $U(3)_A$
relations at all levels ($\eta$-$\eta'$-mixing, $A_1$-pion mixings, pion-photon couplings).
\label{fig:diphoton}}
\end{center}
\end{figure}
At leading order, $C_{\gamma}[\pi_3]=-\frac{1}{f_{\pi}}$, $C_{\gamma}[\eta]=-\frac{1}{\sqrt{3}f_{\pi}}\left(\cos\theta_{\eta}-2\sqrt{2}\sin\theta_{\eta}\right)$ and 
$C_{\gamma}[\eta']=-\frac{1}{\sqrt{3}f_{\pi}}\left(\sin\theta_{\eta}+2\sqrt{2}\cos\theta_{\eta}\right)$. However, we may exploit the experimental measurements of the 
$\pi_3$, $\eta$ and $\eta'$ diphoton widths \cite{Olive:2016xmw} to derive more realistic (though close) estimates: $C_{\gamma}[\pi_3]\simeq-10.75~\mbox{GeV}^{-1}$, $C_{\gamma}[\eta]\simeq-10.8~\mbox{GeV}^{-1}$ and 
$C_{\gamma}[\eta']\simeq-13.6~\mbox{GeV}^{-1}$. While we regard this choice as an educated guess resumming higher-order effects, it could be objected that, in so mixing
orders, cancellations such as those appearing in the $K_L$ diphoton decay amplitude \cite{Gerard:2005yk} are spoilt. Yet, the properties of the CP-odd Higgs -- its mass
or its $\tan\beta$-dependent couplings -- are not so strictly determined by the $U(3)_A$ symmetry as their $K_L$ equivalent, so that we do not expect comparable order-by-order
cancellations. Nevertheless, we will compare our result to the case where strict $U(3)_A$ conditions are enforced, which translates into a larger value of $|\theta_{\eta}|$ together
with the use of leading-order $C_{\gamma}[\pi_3,\eta,\eta']$ as explicited above.
The diphoton width then reads:
\begin{equation}\label{diphotonwidth}
 \Gamma[\tilde{A}_1\to\gamma\gamma]=\frac{\alpha^2m_{A_1}^3}{64\pi^3}\left|O_{AA}\,\left(C_{\gamma}+\delta C_{\gamma}^{e,\mu}\right)+O_{A3}\,C_{\gamma}[\pi_3]+O_{A\eta}\,C_{\gamma}[\eta]+O_{A\eta'}\,C_{\gamma}[\eta']\right|^2
\end{equation}

We show $\Gamma[\tilde{A}_1\to\gamma\gamma]$ in Fig.\ref{fig:diphoton} for\footnote{Most of the amplitudes (or mixing elements) involving $A_1$ depend linearly on $P_{11}$. The only exception is $C_{\gamma}$
where a direct coupling of the higgsinos to the singlet component of $A_1$ intervenes. Although we confine to the case $P_{11}=0.03$ in numerical applications, other choices could thus be easily reconstructed
via a rescaling of the widths by $\sim\left(\frac{P_{11}}{0.03}\right)^2$.} $P_{11}=0.03$, $\tan\beta=10$; the result also depends on the chargino contribution to $C_{\gamma}$: we have 
employed $M_2=\mu_{\mbox{\tiny eff}}=500$~GeV, $\lambda=0.3$. The decay width of Eq.(\ref{diphotonwidth}) corresponds to the (full) red line. The (dashed) blue curve would correspond to a 
pure partonic width, neglecting the mixing with the mesons and including a partonic strange quark with mass $95$~MeV in $C_{\gamma}$. For the (dot-dashed) green curve, we have neglected the partonic
contribution ($C_{\gamma}+\delta C_{\gamma}^{e,\mu}$) and assumed that the width would be purely originating in the meson-mixing. Expectedly, this mixing approximation
provides a qualitatively good agreement with the full result of Eq.(\ref{diphotonwidth}) when $m_{A_1}$ is close to a meson mass. On the other hand, the partonic description 
captures the main effects far from the mixing regime. But it generically falls orders of magnitude away in the vicinity of meson masses. Despite a tiny mixing, we observe that 
the impact of mesons on the diphoton width extends far beyond the immediate vicinity of the meson masses: this is due to
the large mesonic decay widths to photon pairs and the suppressed genuine $A_1$ width for a mostly-singlet state. We note that destructive interferences develop among the various
amplitudes and, in particular, a local cancellation takes place at $m_{A_1}\sim150$~MeV. At $m_{A_1}\sim1.5$~GeV, the impact of the mixing with the $\eta'$ wanes, leaving the 
partonic description in a satisfactory posture again. Finally, the orange dotted curve corresponds to the case where strict $U(3)_A$ relations have been enforced, at the
level of the mixing among pions ($\theta_{\eta}$), between pions and the Higgs state or in the expression of the pion-photon couplings. We observe minor differences, in particular
a small suppression of the $\eta$-driven mixing contribution, but the Higgs width essentially retains the qualitative trends discussed before. The deviation between the red and orange
curves is understood as a higher-order effect and, as such, part of the error estimate. While we stick to the more phenomenological approach below, it is difficult to 
choose which of the two approaches is actually more reliable at this level.

\subsection{Leptonic decays}
The situation is reversed for the leptonic decays: the corresponding branching ratios for the mesons are typically tiny, while the leptonic decays of the Higgs state, $\tan\beta$-enhanced, would naively count 
among the dominant channels at low-mass. Consequently, the muonic and electronic widths of the $\tilde{A}_1$ state are well approximated by their pure-Higgs analogues. On the contrary, the mixing with the mesons
generates an additional leptonic width for the $\pi_3$, $\eta$ and $\eta'$, which represents the main limiting factor on this mixing effect.

\begin{figure}[t]
\begin{center}
\includegraphics[width=0.7\textwidth]{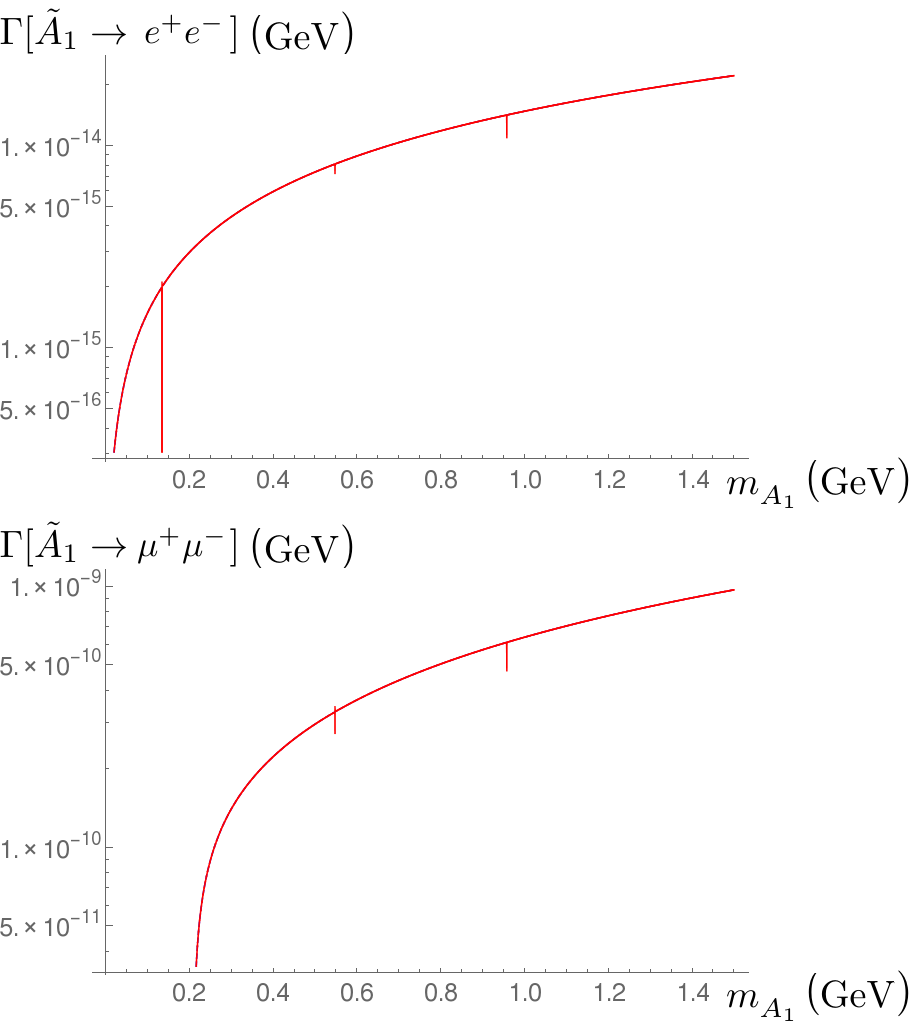}
\caption{Leptonic width of the $\tilde{A}_1$ state. $P_{11}=0.03$, $\tan\beta=10$. The full result (including mixing) is shown as a (full) red line. It essentially covers the (dashed) blue curve,
corresponding to a pure-Higgs decay.
\label{fig:leptonic}}
\end{center}
\end{figure}
We may express the leptonic ($l=e,\mu$) decay width of a pseudoscalar state $P=A_1,\pi_3,\eta,\eta'$ as:
\begin{equation}
 \Gamma[P\to l^+l^-]=\frac{|Y_P^{ll}|^2}{8\pi}m_P\sqrt{1-\frac{4m^2_l}{m_P^2}}\hspace{0.5cm};\hspace{0.5cm}Y_{A_1}^{ll}=\frac{m_l}{\sqrt{2}v}P_{11}\tan\beta
\end{equation}
A few effective mesonic couplings can be estimated numerically from the experimental measurements \cite{Olive:2016xmw}: $Y_{\pi_3}^{ee}\simeq\,3\cdot10^{-7}$;
$Y_{\eta}^{\mu\mu}\simeq\,2\cdot10^{-5}$. However, only upper limits are available for the $e^+e^-$ decays of the $\eta$ and $\eta'$ and the $\mu^+\mu^-$ decay of $\eta'$
is uncharted. We thus neglect such missing input.

We show the leptonic decay widths of the mixed-state $\tilde{A_1}$ in Fig.\ref{fig:leptonic}, for $P_{11}=0.03$, $\tan\beta=10$. We observe that, except for the immediate vicinity of the 
meson masses, the leptonic decays are essentially determined by the pure-Higgs widths.

As we mentioned above, the mixing induces an additional leptonic width for the mesons. This is actually the main impact of the mixing from the perspective of the 
mesons. As the measured leptonic decays are typically small, we may place some limits on this scenario. For instance, still in the case $P_{11}=0.03$, 
$\tan\beta=10$, $\Gamma[\tilde{\pi}^0\to e^+e^-]$ would fall $\sim20\%$ beyond its experimental central value when $m_{A_1}$ is within $\sim0.15$~MeV of $m_{\pi}$. 
Similarly, in a $\sim3$~MeV-wide mass-window centered on $m_{\eta}$, $\Gamma[\tilde{\eta}\to \mu^+\mu^-]$ is $\sim30\%$ off. On the other hand, the limits on
$\Gamma[\tilde{\eta}\to e^+e^-]$ and $\Gamma[\tilde{\eta}'\to e^+e^-]$ are well satisfied. We note, however, that such limits only apply if one assumes that the 
measured $\Gamma[\tilde{\pi}^0\to e^+e^-]$ or $\Gamma[\tilde{\eta}\to \mu^+\mu^-]$ are exactly explained by the SM. When the SM is off, the mixing effect could
well improve the agreement with the measured value. Such a point was actually discussed in \cite{Domingo:2016unq} in the case of $\Gamma[\tilde{\pi}^0\to e^+e^-]$,
as one may choose to see some tension between the experimental measurement \cite{Abouzaid:2006kk} and the theoretical evaluation \cite{Dorokhov:2007bd}.
Thus, exclusion of mass-values for $A_1$ close to a meson mass strongly depends on the assumptions and a detailed analysis would prove necessary.

\subsection{Tri-meson decays}
The decay width of a pseudoscalar $P$ to a tri-mesonic final state $\Pi_i\Pi_j\Pi_k$ may be written as:
\begin{multline}\label{tripion}
 \Gamma[P\to\Pi_i\Pi_j\Pi_k]=\frac{1}{256S_{ijk}\pi^3m_P}\int_{(m_{j}+m_{k})^2}^{(m_P-m_{i})^2}{ds\,\left|{\cal A}_P^{ijk}\right|^2\sqrt{1-\frac{2(m_{j}^2+m_{k}^2)}{s}+\frac{(m_{j}^2-m_{k}^2)^2}{s^2}}}\\{\times\sqrt{\left(1+\frac{s-m_{i}^2}{m_P^2}\right)^2-\frac{4s}{m_P^2}}}
\end{multline}
where $m_{P,i,j,k}$ stand for the masses of $P$, $\Pi_i$, $\Pi_j$, $\Pi_k$ and $S_{ijk}$ is a symmetry factor: $1$, $2$ or $3!$ depending on the number of identical particles in the final state. The transition amplitude ${\cal A}_P^{ijk}$
should be determined from the chiral lagrangian. Expliciting the quartic Higgs-meson couplings in Eq.(\ref{lowen}) provides us with:
\begin{align}\label{quarticA1}
 \tilde{\cal L}_{\chi}&\nonumber\ni-\frac{P_{11}\,A_1}{12\sqrt{2}vf_{\pi}}\cdot\left(\pi_3^2+2\,\pi^+\pi^-\right)\Big\{\left[m_{\pi}^2(\tan^{-1}\beta-\tan\beta)+\delta(\tan^{-1}\beta+\tan\beta)\right]\,\pi_3\\
  &\nonumber+\sqrt{3}\left[m_{\pi}^2(\tan^{-1}\beta+\tan\beta)+\delta(\tan^{-1}\beta-\tan\beta)\right]\left[(\cos\theta_{\eta}-\sqrt{2}\sin\theta_{\eta})\eta+(\sin\theta_{\eta}+\sqrt{2}\cos\theta_{\eta})\eta'\right]\Big\}\\
  &\hspace{0.2cm}-\frac{P_{11}\,A_1}{6\sqrt{2}vf_{\pi}}\cdot\pi_3\,\Big\{\left[m^2_{K^{\pm}}(2\tan^{-1}\beta+\tan\beta)+(m^2_{\pi}-m^2_{K^0})(2\tan^{-1}\beta-\tan\beta)\right]K^+K^-\\
  &\nonumber\hspace{8cm}+\left[m^2_{K^{\pm}}-m^2_{\pi}-3m^2_{K^0}\right]\tan\beta\,K^0\bar{K}^0\Big\}\\
  &\nonumber\hspace{0.2cm}-\frac{P_{11}\,A_1}{6vf_{\pi}}\cdot(K^0K^-\pi^++\bar{K}^0K^+\pi^-)\left[(m^2_{K^{\pm}}+m^2_{\pi})\tan^{-1}\beta-m^2_{K^0}(\tan^{-1}\beta-2\tan\beta)\right]\\
  &\nonumber\hspace{0.2cm}-\frac{P_{11}\,A_1}{12\sqrt{2}vf_{\pi}}\left[m_{\pi}^2(\tan^{-1}\beta-\tan\beta)+\delta(\tan^{-1}\beta+\tan\beta)\right]\,\pi_3\\
  &\nonumber\hspace{6cm}\times\left[(\cos\theta_{\eta}-\sqrt{2}\sin\theta_{\eta})\eta+(\sin\theta_{\eta}+\sqrt{2}\cos\theta_{\eta})\eta'\right]^2
\end{align}
\begin{align}\label{quartic}
 \tilde{\cal L}_{\chi}&\nonumber\ni\frac{m_{\pi}^2}{24f_{\pi}^2}\left(\pi_3^4+4\,\pi_3^2\,\pi^+\pi^-\right)\\
  &\nonumber\hspace{0.2cm}+\frac{\delta}{6\sqrt{3}f_{\pi}^2}\,\left[(\cos\theta_{\eta}-\sqrt{2}\sin\theta_{\eta})\,\eta+(\sin\theta_{\eta}+\sqrt{2}\cos\theta_{\eta})\,\eta'\right]\,\pi_3\,(\pi_3\pi_3+2\pi^+\pi^-)\\
  &\hspace{0.2cm}+\frac{m_{\pi}^2}{12f_{\pi}^2}\left[(\cos\theta_{\eta}-\sqrt{2}\sin\theta_{\eta})\,\eta+(\sin\theta_{\eta}+\sqrt{2}\cos\theta_{\eta})\,\eta'\right]^2\left(\pi_3\pi_3+2\pi^+\pi^-\right)\\
  &\nonumber\hspace{0.2cm}+\frac{1}{12f_{\pi}^2}\Big\{(2m_{K^{\pm}}^2-m^2_{K^0}+m^2_{\pi})\,\pi_3^2K^+K^-+(2m^2_{K^0}-m^2_{K^{\pm}}+m^2_{\pi})\,\pi_3^2K^0\bar{K}^0\\
  &\nonumber\hspace{8cm}+\sqrt{2}\delta\,\pi_3(\pi^+K^-K^0+\pi^-K^+\bar{K}^0)\Big\}\\
  &\nonumber\hspace{0.2cm}+\frac{1}{6\sqrt{3}f_{\pi}^2}\Big\{\left[-3\sqrt{2}\,m^2_{K^{\pm}}\,\sin\theta_{\eta}+(m^2_{\pi}-m^2_{K^0})(\cos\theta_{\eta}-\sqrt{2}\sin\theta_{\eta})\right]\eta\\
  &\nonumber\hspace{1.7cm}+\left[3\sqrt{2}\,m^2_{K^{\pm}}\,\cos\theta_{\eta}+(m^2_{\pi}-m^2_{K^0})(\sin\theta_{\eta}+\sqrt{2}\cos\theta_{\eta})\right]\eta'\Big\}\pi_3K^+K^-\\
  &\nonumber\hspace{0.2cm}+\frac{1}{6\sqrt{3}f_{\pi}^2}\Big\{\left[3\sqrt{2}\,m^2_{K^{0}}\,\sin\theta_{\eta}+(m^2_{K^{\pm}}-m^2_{\pi})(\cos\theta_{\eta}-\sqrt{2}\sin\theta_{\eta})\right]\eta\\
  &\nonumber\hspace{1.7cm}+\left[-3\sqrt{2}\,m^2_{K^{0}}\,\cos\theta_{\eta}-(m^2_{K^{\pm}}-m^2_{\pi})(\sin\theta_{\eta}+\sqrt{2}\cos\theta_{\eta})\right]\eta'\Big\}\pi_3K^0\bar{K}^0\\
  &\nonumber\hspace{0.2cm}+\frac{1}{6\sqrt{6}f_{\pi}^2}\Big\{\left[(2m_{\pi}^2-m^2_{K^{\pm}}-m^2_{K^0})\,\cos\theta_{\eta}-2\sqrt{2}(m^2_{K^0}+m^2_{K^{\pm}}+m^2_{\pi})\sin\theta_{\eta}\right]\eta\\
  &\nonumber\hspace{1.1cm}+\left[(2m_{\pi}^2-m^2_{K^{\pm}}-m^2_{K^0})\,\sin\theta_{\eta}+2\sqrt{2}(m^2_{K^0}+m^2_{K^{\pm}}+m^2_{\pi})\cos\theta_{\eta}\right]\eta'\Big\}(\pi^+K^-K^0+\pi^-K^+\bar{K}^0)\\
  &\nonumber\hspace{0.2cm}+\frac{\delta}{18\sqrt{3}f_{\pi}^2}\pi_3\Big\{(\cos\theta_{\eta}-\sqrt{2}\sin\theta_{\eta})\,\eta+(\sin\theta_{\eta}+\sqrt{2}\cos\theta_{\eta})\,\eta'\Big\}^3
\end{align}

\begin{figure}[htb!]
\begin{center}
\includegraphics[width=0.67\textwidth]{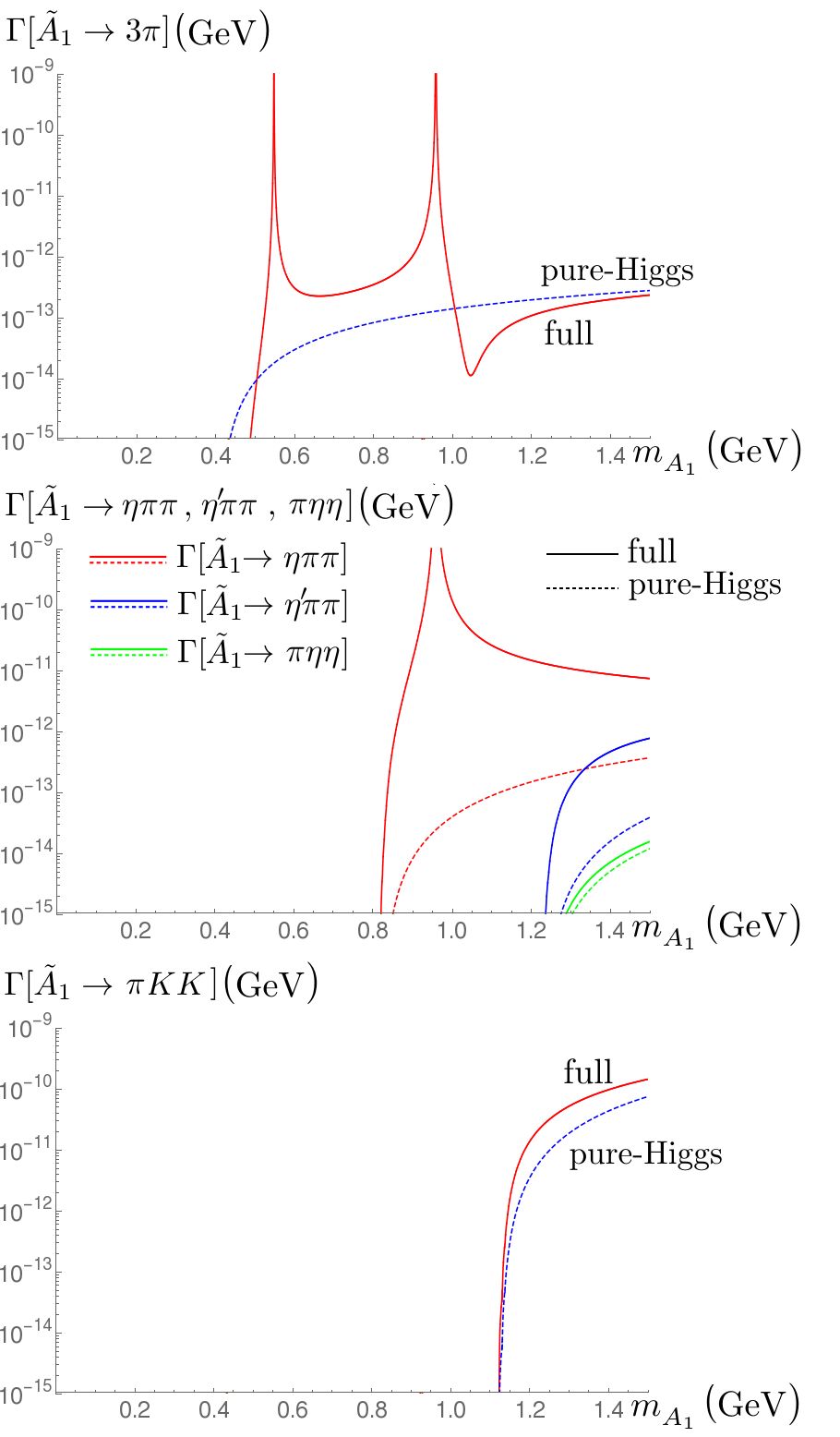}
\caption{Hadronic decay widths to for $P_{11}=0.03$, $\tan\beta=10$. The plot above corresponds to the $\tilde{A}_1\to3\pi$ channel: the full result (including mixing) is shown as a (full) red line while the (dashed) blue curve
corresponds to a pure-Higgs decay. The plot in the middle shows the $\tilde{A}_1\to\eta\pi\pi$ (red curves: full $\to$ with mixing; dashed $\to$ without mixing), $\tilde{A}_1\to\eta'\pi\pi$ (blue curves)
and $\tilde{A}_1\to\pi\eta\eta$ (green curves) widths. The plot on the bottom corresponds to the $\tilde{A}_1\to\pi KK$ channels (red full curve: with mixing; blue dashed cure: without).
\label{fig:3Pi}}
\end{center}
\end{figure}
Starting from $m_{A_1}\gsim3m_{\pi}$, the tri-pion decays of $A_1$ become kinematically accessible. The chiral lagrangian contains $A_1$-pion couplings explicitly,
$\propto\frac{m_{\pi}^2}{vf_{\pi}}P_{11}(\tan^{-1}\beta-\tan\beta)$. The $\eta$ and $\eta'$ couplings to three pions are isospin-violating ($\propto\delta$). 
Diagrams involving the $\eta/\eta'$-$\pi_3$ mixing thus contribute at the same order in $\delta$ \cite{Cronin:1967jq,Osborn:1970nn}. The basic evaluation employing $\delta\simeq 
m^2_{K^{\pm}}-m^2_{K^0}-m^2_{\pi^{\pm}}+m^2_{\pi}$ is known to provide a substantially smaller contribution to $\Gamma[\eta\to3\pi]$ than experiments indicate \cite{Holstein:2001bt}. 
In practice we thus rescale the corresponding couplings by comparison with the experimental widths $\Gamma[\eta\to3\pi]$ and $\Gamma[\eta'\to3\pi]$ directly. We show
the corresponding estimate of $\Gamma[\tilde{A}_1\to3\pi]$ (summing over neutral and charged pionic final states) for $P_{11}=0.03$, $\tan\beta=10$ in the upper part of Fig.\ref{fig:3Pi}. 
The full result (red full curve) differs again sizably from the pure-Higgs width (blue dashed line) due to the large impact of the $\eta$ and $\eta'$ resonances.

The $\eta\pi\pi$ channels open up at $m_{A_1}\gsim820$~MeV. In the case of the $\eta'$, they represent one of the main final states -- the $\eta'\eta\pi\pi$ coupling 
is isospin-conserving $\propto \frac{m^2_{\pi}}{f^2_{\pi}}$, though the decay is phase-space suppressed. 
Again, we rescale the couplings of Eq.(\ref{quartic}) in order to account for the experimental $\Gamma[\eta'\to\eta\pi\pi]$. Similarly, we include the $\eta'\pi\pi$
and $\pi_3\eta\eta$ channels at $m_{A_1}\gsim1.23$~GeV. It is remarkable that in none of these decays the coupling $\propto \frac{Bm_s}{f_{\pi}}$ intervenes. The results are displayed in the plot
in the middle of Fig.\ref{fig:3Pi}: $\tilde{A}_1\to\eta\pi\pi$ (red lines) expectedly proves the most relevant of these channels. The very large $\eta'\to\eta\pi\pi$ decay
induces a sizable deviation of the $\tilde{A}_1$ decay (full curve) as compared to the pure-Higgs amplitude (dashed curve) and this effect is still partially affecting the $\tilde{A}_1\to\eta\pi\pi$
width at $m_{A_1}\sim1.5$~GeV.

Beyond $m_{A_1}\gsim1.12$~GeV, the $\pi KK$ channels are accessible in their turn. They are the first decays to employ the coupling $\propto \frac{Bm_s}{f_{\pi}}$, meaning that the 
impact of the strange quark on the $\tilde{A_1}$ width is kinematically delayed till this quite-high threshold. We cannot use experimental data to evaluate the $\pi_3$, $\eta$ or $\eta'$ couplings
to the corresponding final states in a phenomenologically more efficient way than employing Eq.(\ref{quartic}). However, we note that the kinematically relevant 
region is already far above the masses of the pseudoscalar mesons, so that the mixing effect should be subdominant. Our result (summing over the $\pi_3K^+K^-$, 
$\pi_3K^0\bar{K}^0$, $\pi^+K^-K^0$ and $\pi^-K^+\bar{K}^0$ final states) is displayed in the lower plot of Fig.\ref{fig:3Pi}. These kaonic widths are typically one to
two orders of magnitude larger than the pionic decays, due to the larger coupling. The mixing effect appears to affect these decay channels in a subdominant
way, although a small excess is still visible at $m_{A_1}\sim1.5$~GeV.

Other tri-meson final states include $\pi_ 3\eta\eta'$, $\pi_3\eta'\eta'$, $3\eta$, $\eta'\eta\eta$, $\eta\eta'\eta'$, $3\eta'$, $\eta K^+K^-$, $\eta'K^+K^-$, 
$\eta K^0\bar{K}^0$ and $\eta' K^0\bar{K}^0$. They intervene only beyond $m_{A_1}>1.5$~GeV so we leave the corresponding description to the following section.

\subsection{Radiative hadronic decays}
The decays $\eta\to\gamma\pi^+\pi^-$ and $\eta'\,(\to\gamma\,\rho,\omega)\to\gamma\pi^+\pi^-$ show the relevance of radiative decay modes for light pseudoscalar states. Such decays are
entirely specified by the anomaly and (in the case of pseudoscalar mesons) they are well described in a Vector Dominance approach -- see e.g. \cite{Holstein:2001bt}. 
For the Higgs pseudoscalar, we confine to the leading-order, in which the radiative hadronic decays result from the mixing effect with the pseudoscalar mesons.

\begin{figure}[t]
\begin{center}
\includegraphics[width=0.67\textwidth]{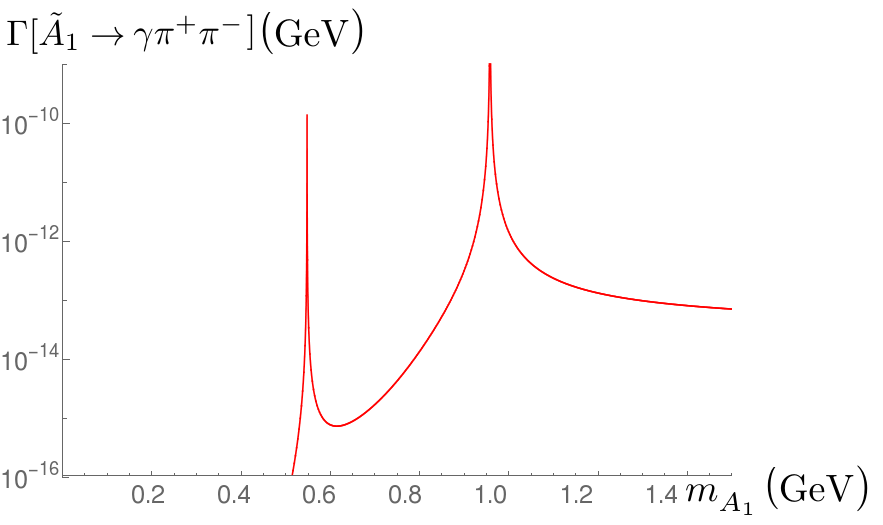}
\caption{$\tilde{A}_1\to\gamma\pi^+\pi^-$ width for $P_{11}=0.03$, $\tan\beta=10$. In our approach, this decay is only mediated by the mixing with the pseudoscalar mesons.
\label{fig:Radhadr}}
\end{center}
\end{figure}
We follow the description of \cite{Holstein:2001bt} for the $\eta/\eta'\to\gamma\pi^+\pi^-$ decay and correct the small mismatch with the experimental widths by a
rescaling factor. Then we display the decay width acquired by $\tilde{A}_1$ via mixing in Fig.\ref{fig:Radhadr}. Expectedly, the effect is largest close to the 
$\eta$ and $\eta'$ masses. Yet, the tail at $m_{A_1}\simeq1.5$~GeV competes in magnitude with the pionic decay widths. We may 
wonder whether large $\gamma KK$ decays could not develop via the mediation of the $\phi$. We will assume that it is not the case, first because the $\phi$ is much 
narrower than the $\rho$, second because we believe that these radiative decays remain subdominant as compared to the $\pi KK$ channels.

\subsection{Decays in the chiral limit: summary}
\begin{figure}[tb]
\begin{center}
\includegraphics[width=0.67\textwidth]{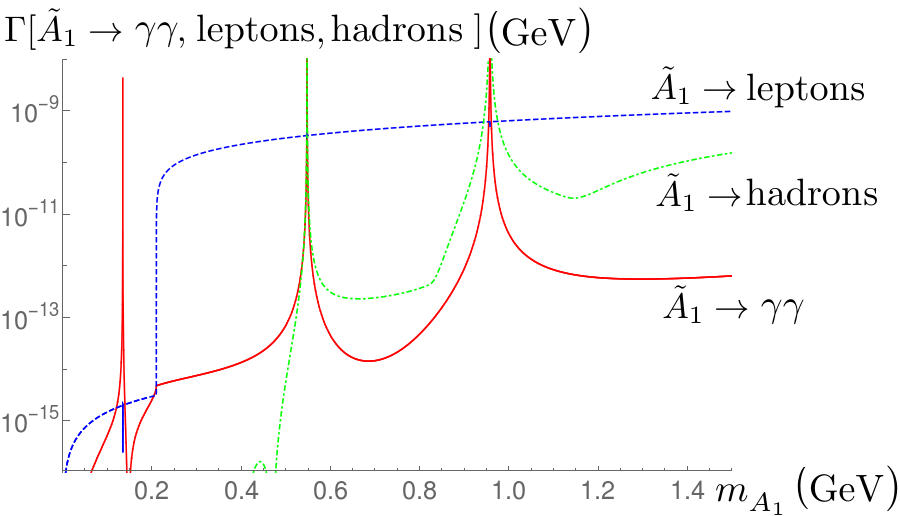}
\caption{Summary of the $\tilde{A}_1$ decays in the chiral limit for $P_{11}=0.03$, $\tan\beta=10$: the full red line corresponds to the diphoton decay, the blue dashed curve, to the leptonic decays, 
and the green dot-dashed curve sums all hadronic decays.
\label{fig:chiral}}
\end{center}
\end{figure}

At this level, we believe to have considered the major possible decay channels to SM particles for the CP-odd Higgs in the chiral limit. These are summarized in
Fig.\ref{fig:chiral}: as was already pointed out by e.g.\ \cite{Dolan:2014ska}, the leptonic widths (blue dashed curve) dominate most of the low-mass region. One 
then naively expects severe constraints from e.g.\ flavour observables, where, however, the interplay of supersymmetric contributions in the flavour-changing $A_1$
couplings should be considered carefully: such a discussion goes beyond the aims of the present paper.
However, close to the mass of the $\pi_3$, $\eta$ or $\eta'$, the Higgs pseudoscalar may have enhanced decays to a photon pair (red full curve) or to hadrons 
(green dot-dashed curve) and these final states may compete with the dimuon channel. At $m_{A_1}\simeq1.5$~GeV, the hadronic decays still represent only $\sim10\%$
of the SM width of $\tilde{A}_1$ (for the particular values  $P_{11}=0.03$, $\tan\beta=10$).

As singlino Dark Matter is a motivation for the light $A_1$ scenario, we note that an invisible decay into singlinos ($\tilde{\chi}_s^0$) could be relevant if the $A_1$ mass is above threshold.
We expect this channel to be dominated by the singlet-singlino interactions. At leading order:
\begin{equation}\label{singdec}
 \Gamma[\tilde{A}_1\to\tilde{\chi}_s^0\tilde{\chi}_s^0]=\frac{\kappa^2(1-P_{11}^2)}{8\pi}m_{A_1}\sqrt{1-\frac{4m_{\tilde{\chi}_s^0}}{m_{A_1}}}
\end{equation}
For the thermal annihilation of singlinos in the early universe to be efficient enough -- so that we avoid excessive relics --, the corresponding cross-section must be enhanced by a resonant $A_1$, i.e.
$m_{A_1}$ should be close to the energy of the singlino pair maximizing the Boltzmann distribution at the freeze-out temperature, i.e\ $m_{A_1}\simeq2m_{\tilde{\chi}_s^0}$. Then, the phase-space of the 
decay is suppressed. Moreover, observing that the singlino mass in the $Z_3$-conserving NMSSM is given by $m_{\tilde{\chi}_s^0}\simeq2\kappa s$, it seems
natural to turn to the Peccei-Quinn limit ($\kappa\to0$) to ensure both a light $A_1$ and a light singlino. Eq.(\ref{singdec}) is further suppressed in this limit.
However, in view of $m_{A_1}\simeq2m_{\tilde{\chi}_s^0}$, threshold corrections would also be relevant. We will not enter into more details here
as we are chiefly interested in the SM decays of the CP-odd Higgs. We remind that sizable invisible $A_1$ decays could fall under direct limits from $K$ or $B$ decays, 
since hadronic processes may be converted to hadronic+invisible, which receive tight phenomenological constraints. Similarly to the case of large leptonic decays, 
the impact of such limits on the details of the supersymmetric spectrum should be studied carefully, which is beyond the scope of the present paper.

\section{Beyond the chiral limit}
In the previous section, we have seen how the interactions of a light CP-odd Higgs with the strong-interacting sector could be described in a chiral lagrangian. The mixing of $A_1$ with the neutral
pseudoscalar mesons appeared as an important ingredient, coupling the Higgs pseudoscalar to the chiral anomaly and modifying its hadronic decays. Yet, beyond
$m_{A_1}\gsim1$~GeV, one becomes suspicious as to the validity of the chiral description and one prefers to turn to a more partonic approach, the 
perturbative spectator model \cite{McKeen:2008gd,Dolan:2014ska,Gunion:1989we}. This effective approach essentially adopts a partonic dynamics while keeping the 
kinematics of the hadrons. In this section, we re-cast the hadronic decays of the Higgs pseudoscalar in terms of this partonic description.

We thus consider the following effective lagrangian for the interaction of $A_1$ with the partons:
\begin{equation}\label{specmod}
  {\cal L}_{\mbox{\tiny spect.}}=\frac{\imath}{\sqrt{2}}A_1\left\{{\cal Y}^A_u\,\bar{u}\gamma_5u+{\cal Y}^A_d\,\bar{d}\gamma_5d+{\cal Y}^A_s\,\bar{s}\gamma_5s\right\}
\end{equation}
${\cal Y}^A_u$, ${\cal Y}^A_d$ and ${\cal Y}^A_s$ are effective Yukawa couplings that should be identified with the chiral couplings in the chiral limit. The 
partonic amplitudes, stripped of the spinors (since these affect the kinematics), are particularly simple and read:
\begin{align}
 &\tilde{\cal A}[A_1\to u\bar{u}]=\frac{1}{\sqrt{2}}{\cal Y}^A_u\\
 &\tilde{\cal A}[A_1\to d\bar{d}]=\frac{1}{\sqrt{2}}{\cal Y}^A_d\nonumber\\
 &\tilde{\cal A}[A_1\to s\bar{s}]=\frac{1}{\sqrt{2}}{\cal Y}^A_s\nonumber
\end{align}
Our concern now consists in distributing this dynamics among the hadronic channels. We shall assume that these are dominated by the tri-meson final states.

Forgetting momentarily about the mixing effect (i.e.\ we focus on genuine Higgs amplitudes below), it is useful to notice that the chiral amplitudes of 
Eq.(\ref{tripion}) satisfy the property:
\begin{multline}\label{sumrule}
 \sum_{(i,j,k)}\frac{1}{S_{ijk}}\left|{\cal A}_P^{ijk}\right|^2=18\left(\frac{BP_{11}}{6vf_{\pi}^2}\right)^2\left[\left(\frac{m_u}{\tan\beta}\right)^2+\left(m_d\tan\beta\right)^2+\left(m_s\tan\beta\right)^2\right.\\+2m_um_d+2m_um_s+2m_dm_s\,\tan^2\beta\Big]
\end{multline}
If we discard the terms of the second line, subleading in $m_s$ or in $\tan\beta$ in individual amplitudes, we may identify Eq.(\ref{sumrule}) with its partonic
analogue $\frac{N_c}{2}\left[\left({\cal Y}^A_u\right)^2+\left({\cal Y}^A_d\right)^2+\left({\cal Y}^A_s\right)^2\right]$ and come to the relations:
\begin{equation}\label{matching}
 {\cal Y}^A_u\simeq\frac{BP_{11}}{\sqrt{3}vf_{\pi}^2}\frac{m_u}{\tan\beta}\hspace{0.5cm};\hspace{0.5cm}{\cal Y}^A_d\simeq\frac{BP_{11}}{\sqrt{3}vf_{\pi}^2}\,m_d\tan\beta\hspace{0.5cm};\hspace{0.5cm}{\cal Y}^A_s\simeq\frac{BP_{11}}{\sqrt{3}vf_{\pi}^2}\,m_s\tan\beta
\end{equation}
Eq.(\ref{sumrule}) also hints at how to distribute the partonic amplitude among the 21 tri-meson final states:
\begin{itemize}
 \item $\tilde{\cal A}^2[A_1\to 3\pi]=\frac{5}{144}N_c\left[\left({\cal Y}^A_u\right)^2+\left({\cal Y}^A_d\right)^2\right]$;
 \item $\tilde{\cal A}^2[A_1\to \eta\pi\pi]=\frac{1}{16}N_c\left[\left({\cal Y}^A_u\right)^2+\left({\cal Y}^A_d\right)^2\right]\left(\cos\theta_{\eta}-\sqrt{2}\sin\theta_{\eta}\right)^2$;
 \item $\tilde{\cal A}^2[A_1\to \eta'\pi\pi]=\frac{1}{16}N_c\left[\left({\cal Y}^A_u\right)^2+\left({\cal Y}^A_d\right)^2\right]\left(\sin\theta_{\eta}+\sqrt{2}\cos\theta_{\eta}\right)^2$;
 \item $\tilde{\cal A}^2[A_1\to \pi\eta\eta]=\frac{1}{144}N_c\left[\left({\cal Y}^A_u\right)^2+\left({\cal Y}^A_d\right)^2\right]\left(\cos\theta_{\eta}-\sqrt{2}\sin\theta_{\eta}\right)^4$;
 \item $\tilde{\cal A}^2[A_1\to \pi\eta\eta']=\frac{1}{72}N_c\left[\left({\cal Y}^A_u\right)^2+\left({\cal Y}^A_d\right)^2\right]\left(\cos\theta_{\eta}-\sqrt{2}\sin\theta_{\eta}\right)^2\left(\sin\theta_{\eta}+\sqrt{2}\cos\theta_{\eta}\right)^2$;
 \item $\tilde{\cal A}^2[A_1\to \pi\eta'\eta']=\frac{1}{144}N_c\left[\left({\cal Y}^A_u\right)^2+\left({\cal Y}^A_d\right)^2\right]\left(\sin\theta_{\eta}+\sqrt{2}\cos\theta_{\eta}\right)^4$;
 \item $\tilde{\cal A}^2[A_1\to 3\eta]=\frac{1}{1296}N_c\left[\left({\cal Y}^A_u\right)^2+\left({\cal Y}^A_d\right)^2+64\left({\cal Y}^A_s\right)^2\right]$ (in the approximation $\theta_{\eta}\simeq0$);
 \item $\tilde{\cal A}^2[A_1\to \eta\eta\eta']=\frac{1}{216}N_c\left[\left({\cal Y}^A_u\right)^2+\left({\cal Y}^A_d\right)^2+16\left({\cal Y}^A_s\right)^2\right]$ (in the approximation $\theta_{\eta}\simeq0$);
 \item $\tilde{\cal A}^2[A_1\to \eta\eta'\eta']=\frac{1}{108}N_c\left[\left({\cal Y}^A_u\right)^2+\left({\cal Y}^A_d\right)^2+4\left({\cal Y}^A_s\right)^2\right]$ (in the approximation $\theta_{\eta}\simeq0$);
 \item $\tilde{\cal A}^2[A_1\to 3\eta']=\frac{1}{162}N_c\left[\left({\cal Y}^A_u\right)^2+\left({\cal Y}^A_d\right)^2+\left({\cal Y}^A_s\right)^2\right]$ (in the approximation $\theta_{\eta}\simeq0$);
 \item $\tilde{\cal A}^2[A_1\to \pi KK]=\frac{1}{36}N_c\left[4\left({\cal Y}^A_u\right)^2+4\left({\cal Y}^A_d\right)^2+3\left({\cal Y}^A_s\right)^2\right]$;
 \item $\tilde{\cal A}^2[A_1\to \eta KK]=\frac{1}{12}N_c\left[\left({\cal Y}^A_s\right)^2\right]$ (in the approximation $\theta_{\eta}\simeq0$);
 \item $\tilde{\cal A}^2[A_1\to \eta' KK]=\frac{1}{12}N_c\left[\left({\cal Y}^A_u\right)^2+\left({\cal Y}^A_d\right)^2+2\left({\cal Y}^A_s\right)^2\right]$ (in the approximation $\theta_{\eta}\simeq0$).
\end{itemize}
\begin{figure}[tbh!]
\begin{center}
\includegraphics[width=0.67\textwidth]{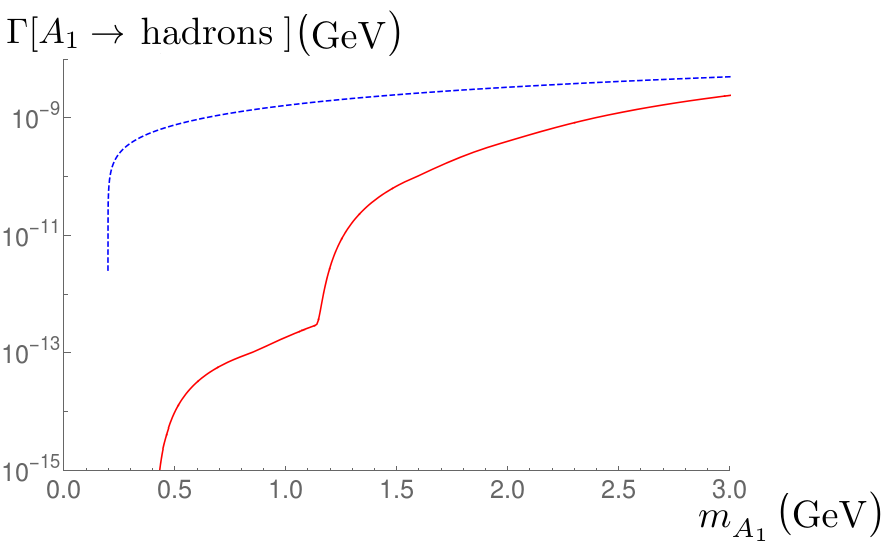}
\caption{Pure Higgs decays in the spectator approach for $P_{11}=0.03$, $\tan\beta=10$: the full red line sums all tri-meson channels. The blue dashed curve corresponds
to the two-body decays into quarks obtained for perturbative-quark masses of $m_u=2$~MeV, $m_d=4$~MeV and $m_s=95$~MeV.
\label{fig:spec}}
\end{center}
\end{figure}
We may now determine ${\cal Y}^A_u$, ${\cal Y}^A_d$ and ${\cal Y}^A_s$ from a formal matching of the $A_1\to 3\pi$ and $A_1\to\pi KK$ amplitudes in the chiral
and the spectator approaches, which returns the expressions of Eq.(\ref{matching}). From this distribution, we can derive the picture of Fig.\ref{fig:spec}, where the 
sum of the tri-meson decay-widths is shown as a full red curve.  It essentially appears as the superposition of two decay widths with respective thresholds $\sim0.4$~GeV (for 
the light $u$, $d$ quarks) and $\sim1.1$~GeV (for the $s$ quark). Interestingly, this total tri-meson width appears to converge slowly towards the quark-partonic width
(blue dashed curve) obtained for perturbative-quark masses of $m_u=2$~MeV, $m_d=4$~MeV and $m_s=95$~MeV. A transition to the perturbative quark-regime may thus become 
relevant around $m_{A_1}\gsim3$~GeV. From this perspective, the impact of mesons for the $A_1$ appears as a delayed kinematic opening of the quark decays.

\begin{figure}[tb]
\begin{center}
\includegraphics[width=0.67\textwidth]{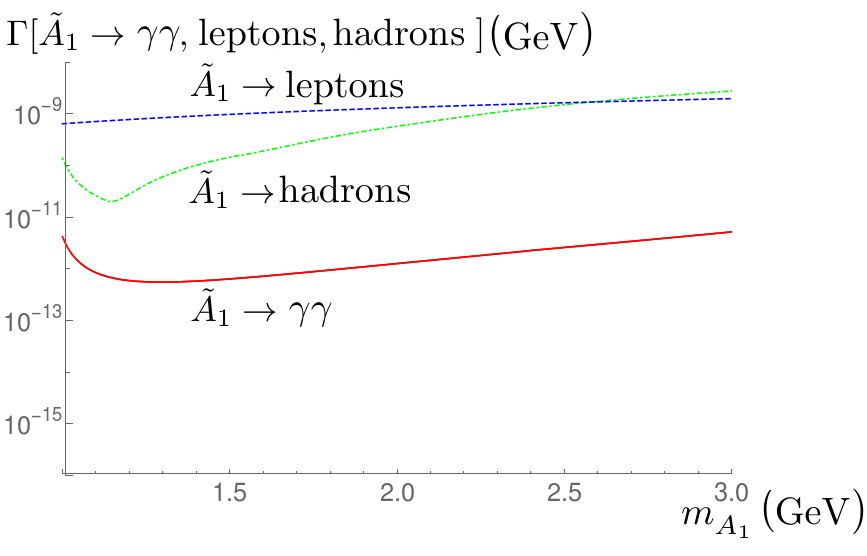}
\caption{Higgs decays in the spectator approach (including meson-mixing) for $P_{11}=0.03$, $\tan\beta=10$: the green dot-dashed line sums all tri-meson channels. The 
blue dashed curve corresponds to the leptonic decays and the full red curve to the diphoton width.
\label{fig:specwidth}}
\end{center}
\end{figure}

We had temporarily forgotten about the $A_1$-meson mixing. This effect is present in Eq.(\ref{specmod}) however, if we evaluate the partonic operators for the 
meson wave functions. Below, we shall keep the coefficients derived in the chiral limit. It is understood that the mixing effect should disappear slowly as 
$m_{A_1}$ is farther away from the $\eta'$ mass. Yet, we have seen in the chiral limit that the sizable $\eta/\eta'$ couplings to mesons may extend some influence up
to $m_{A_1}\sim1.5$~GeV. Adding this ingredient to the spectator widths, we arrive at the picture of Fig.\ref{fig:specwidth}. The diphoton (full red line) and the 
hadronic (green dot-dashed) widths are still under the influence of the $\eta/\eta'$ at $m_{A_1}\simeq1$~GeV but eventually converge towards a decoupled regime at 
$m_{A_1}\simeq3$~GeV. We also observe that the hadronic width eventually becomes competitive with the leptonic one (blue dashed curve) around $m_{A_1}\simeq2.5-3$~GeV.

Close to $m_{A_1}\simeq3$~GeV, the charm threshold opens up. The impact on the decays of the CP-odd Higgs is not necessarily large, as the $A_1c\bar{c}$ coupling is
$\tan\beta$-suppressed. As for the interaction with the lighter quarks -- or at the $b\bar{b}$ threshold \cite{Domingo:2011rn} --, the first effect that can be expected
is a mixing of the CP-odd Higgs with the pseudoscalar charmonia $\eta_c(nS)$ -- in particular $m_{\eta_c(1S)}\simeq2.98$~GeV. Then genuine $c\bar{c}$ decays become 
kinematically allowed when $m_{A_1}\gsim m_{\pi}+2m_{D}\simeq3.9$~GeV. In the meanwhile, however, the $\tau^+\tau^-$ threshold has been reached and the $A_1\to\tau^+\tau^-$
decay, $\tan^2\beta$-enhanced, should dominate the disintegrations of the pseudoscalar Higgs, placing all the hadronic or muonic branching ratios at the percent level.
As our focus in this paper is the very-low mass region, we will not detail these effects here.

\begin{figure}[tb]
\begin{center}
\includegraphics[width=1\textwidth]{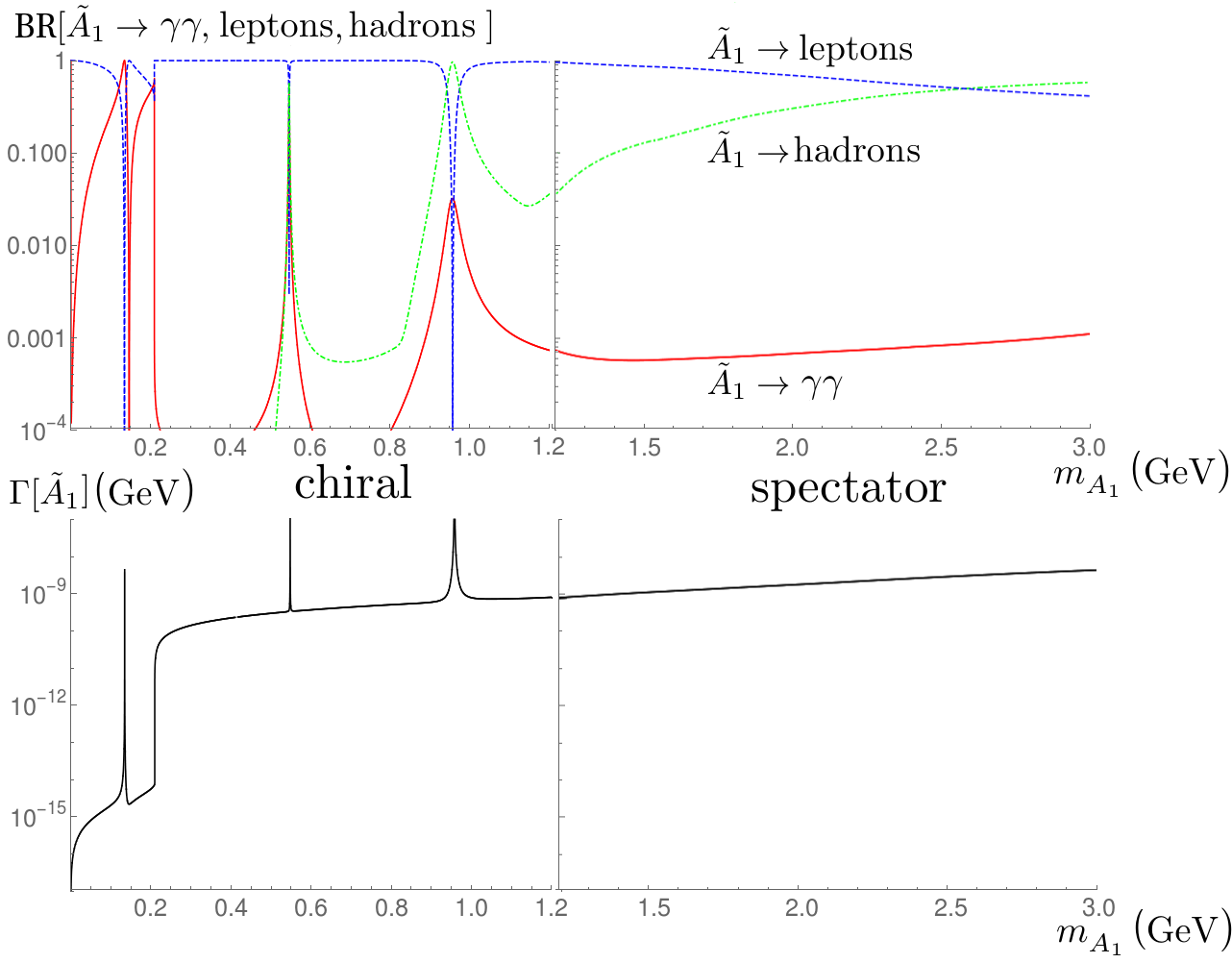}
\caption{Upper plot: Higgs branching ratios for $P_{11}=0.03$, $\tan\beta=10$: hadronic (green dot-dashed), leptonic (blue dashed), diphoton (red full). Lower plot: total $A_1$ width.
The chiral model has been used up to $m_{A_1}=1.2$~GeV. Beyond, we employ the spectator approach.
\label{fig:totalwidth}}
\end{center}
\end{figure}

Before closing this discussion, we display the branching ratios of the light pseudoscalar for $P_{11}=0.03$, $\tan\beta=10$ in Fig.\ref{fig:totalwidth}, both in the chiral and the 
spectator approaches. We assume that there is no invisible decay. Expectedly, the leptonic decays (blue dashed curve) dominate over a wide range of mass. However, 
the diphoton channel can be competitive at low mass, in the vicinity of $m_{A_1}\simeq m_{\pi}$ or just below the dimuon threshold. The hadronic decays become sizable
at $m_{A_1}\simeq3$~GeV or close to the $\eta$ and $\eta'$ masses. The total width is shown in the lower plot of Fig.\ref{fig:totalwidth}. The general scale is that of
the dimuon decay width, but the meson resonances are visible as small spikes. Below the dimuon threshold and with the exception of $m_{A_1}\sim m_{\pi}$, the CP-odd Higgs is relatively long-lived.
Considering the boost factor of order (at least) $\sim100$ in LHC searches for $H[125]\to2A_1$, $A_1$ would fly centimeters before decaying, leading to displaced vertices. 
With even larger boost factors (due to e.g.\ lower $m_{A_1}$, sizable longitudinal energy) or smaller $P_{11}$, the pseudoscalar may well escape the detectors, thus appearing as missing energy. 
Above the dimuon treshold, the CP-odd Higgs is reasonably short-lived and should decay within $\mu$m, unless $P_{11}$ is extremely small.

\vspace{1cm}

It is now time to summarize our achievements in this paper. We have shown how the interactions of a light NMSSM Higgs pseudoscalar with the hadronic sector could be described
by a chiral lagrangian. A remarkable effect, which, to our knowledge, has not received much attention before, appears in the form of a mixing of the CP-odd Higgs with the 
pseudoscalar mesons. This mixing has little effect on the mesons themselves, since they have large hadronic or anomaly-driven decays. On the other hand, its impact on the 
naturally narrow (singlet-dominated) $A_1$ can be very important. In particular, the photonic and hadronic decay widths are sizably affected. However, the leptonic
final state remains the dominant decay channel of the Higgs state in most of the considered mass-range. As the mesons do not possess large decays into leptons, it is possible
to place limits on the leptonic width that they acquire via mixing. The chiral lagrangian also provides quartic $A_1$-meson interaction terms. These govern the tri-meson
decays of the pure-Higgs component. Beyond $m_{A_1}\simeq1$~GeV, it is possible to extend the description of the Higgs-interactions with the strong-interacting sector
using the spectator approach.

Our results can be criticized in many ways. In particular, sizable higher order corrections are known to modify the chiral couplings of the mesons, so that it seems questionable
to rely on the tree-level chiral vertices for the Higgs state. A more comprehensive approach, using higher-dimensional chiral operators as well as loop effects, has 
proved (partially) successful in describing the hadronic decays of the $\eta$ and $\eta'$ and the same type of exercise could probably be adapted with profit to
the case of the light CP-odd Higgs, increasing the reliability of the computed widths. Yet, such a calculation is far beyond the ambitions of this paper where our
scope is limited to a qualitative investigation of the relevant effects. Consequently, even though the picture that we propose is somewhat more convincing than
a pure partonic, quark-perturbative approach, we should remain aware that large corrections (of order $100\%$) could modify the actual magnitude of the hadronic decays.

Deficient they are, our estimates of the pseudoscalar Higgs widths can be applied to constrain the rich phenomenology of this particle. In particular, limits
from flavour-physics, quarkonium decays or beam-dump experiment should be considered in this new light. Still, an additional decay channel should be considered in 
the presence of a very-light Dark Matter candidate. A Fortran implementation of the $A_1$-decays in the low-mass region is in preparation at the request of 
\verb|NMSSMTools| \cite{NMSSMTools}.

Finally, we remind that the effects that we have described here in the explicit case of the NMSSM could be easily transposed to other models including a light
CP-odd state.

\section*{Acknowledgements}
The author thanks U.~Ellwanger for comments. This work has been supported by CICYT (grant FPA 2013-40715-P).


\begin{thebibliography}{100}
 
\bibitem{Ellwanger:2009dp}
  U.~Ellwanger, C.~Hugonie and A.~M.~Teixeira,
  Phys.\ Rept.\  {\bf 496} (2010) 1
  [arXiv:0910.1785 [hep-ph]].

\bibitem{Bechtle:2016kui}
  P.~Bechtle, H.~E.~Haber, S.~Heinemeyer, O.~Stål, T.~Stefaniak, G.~Weiglein and L.~Zeune,
  arXiv:1608.00638 [hep-ph].
  
\bibitem{Domingo:2015eea}
  F.~Domingo and G.~Weiglein,
  JHEP {\bf 1604} (2016) 095
  [arXiv:1509.07283 [hep-ph]].
  
\bibitem{Barate:2003sz}
  R.~Barate {\it et al.} [LEP Working Group for Higgs boson searches and ALEPH and DELPHI and L3 and OPAL Collaborations],
  Phys.\ Lett.\ B {\bf 565} (2003) 61
  [hep-ex/0306033].

\bibitem{Aad:2014kga}
  G.~Aad {\it et al.} [ATLAS Collaboration],
  JHEP {\bf 1503} (2015) 088
  [arXiv:1412.6663 [hep-ex]].
  
\bibitem{Khachatryan:2015qxa}
  V.~Khachatryan {\it et al.} [CMS Collaboration],
  JHEP {\bf 1511} (2015) 018
  [arXiv:1508.07774 [hep-ex]].

\bibitem{Dobrescu:2000jt}
  B.~A.~Dobrescu, G.~L.~Landsberg and K.~T.~Matchev,
  Phys.\ Rev.\ D {\bf 63} (2001) 075003
  [hep-ph/0005308].

\bibitem{Dobrescu:2000yn}
  B.~A.~Dobrescu and K.~T.~Matchev,
  JHEP {\bf 0009} (2000) 031
  [hep-ph/0008192].
  
\bibitem{Dermisek:2006wr}
  R.~Dermisek and J.~F.~Gunion,
  Phys.\ Rev.\ D {\bf 75} (2007) 075019
  [hep-ph/0611142].
  
\bibitem{Morrissey:2008gm}
  D.~E.~Morrissey and A.~Pierce,
  Phys.\ Rev.\ D {\bf 78} (2008) 075029
  [arXiv:0807.2259 [hep-ph]].
  
\bibitem{Hiller:2004ii}
  G.~Hiller,
  Phys.\ Rev.\ D {\bf 70} (2004) 034018
  [hep-ph/0404220].

\bibitem{Domingo:2007dx}
  F.~Domingo and U.~Ellwanger,
  JHEP {\bf 0712} (2007) 090
  [arXiv:0710.3714 [hep-ph]].
  
\bibitem{Heng:2008rc}
  Z.~Heng, R.~J.~Oakes, W.~Wang, Z.~Xiong and J.~M.~Yang,
  Phys.\ Rev.\ D {\bf 77} (2008) 095012
  [arXiv:0801.1169 [hep-ph]].
  
\bibitem{Andreas:2010ms}
  S.~Andreas, O.~Lebedev, S.~Ramos-Sanchez and A.~Ringwald,
  JHEP {\bf 1008} (2010) 003
  [arXiv:1005.3978 [hep-ph]].
  
\bibitem{Domingo:2015wyn}
  F.~Domingo,
  Eur.\ Phys.\ J.\ C {\bf 76} (2016) no.8,  452
  [arXiv:1512.02091 [hep-ph]].
  
\bibitem{Drees:1989du}
  M.~Drees and K.~i.~Hikasa,
  Phys.\ Rev.\ D {\bf 41} (1990) 1547.
  
\bibitem{SanchisLozano:2002pm}
  M.~A.~Sanchis-Lozano,
  Mod.\ Phys.\ Lett.\ A {\bf 17} (2002) 2265
  [hep-ph/0206156].
  
\bibitem{SanchisLozano:2003ha}
  M.~A.~Sanchis-Lozano,
  Int.\ J.\ Mod.\ Phys.\ A {\bf 19} (2004) 2183
  [hep-ph/0307313].
  
\bibitem{SanchisLozano:2005di}
  M.~A.~Sanchis-Lozano,
  PoS HEP {\bf 2005} (2006) 334
  [hep-ph/0510374].
  
\bibitem{McElrath:2005bp}
  B.~McElrath,
  Phys.\ Rev.\ D {\bf 72} (2005) 103508
  [hep-ph/0506151].
  
\bibitem{SanchisLozano:2006gx}
  M.~A.~Sanchis-Lozano,
  J.\ Phys.\ Soc.\ Jap.\  {\bf 76} (2007) 044101
  [hep-ph/0610046].
  
\bibitem{Dermisek:2006py}
  R.~Dermisek, J.~F.~Gunion and B.~McElrath,
  Phys.\ Rev.\ D {\bf 76} (2007) 051105
  [hep-ph/0612031].
  
\bibitem{Fullana:2007uq}
  E.~Fullana and M.~A.~Sanchis-Lozano,
  Phys.\ Lett.\ B {\bf 653} (2007) 67
  [hep-ph/0702190].

\bibitem{Hodgkinson:2008ei}
  R.~N.~Hodgkinson,
  Phys.\ Lett.\ B {\bf 665} (2008) 219
  [arXiv:0802.3197 [hep-ph]].

\bibitem{Domingo:2008rr}
  F.~Domingo, U.~Ellwanger, E.~Fullana, C.~Hugonie and M.~A.~Sanchis-Lozano,
  JHEP {\bf 0901} (2009) 061
  [arXiv:0810.4736 [hep-ph]].
  
\bibitem{McKeen:2008gd}
  D.~McKeen,
  Phys.\ Rev.\ D {\bf 79} (2009) 015007
  [arXiv:0809.4787 [hep-ph]].
  
\bibitem{Domingo:2009tb}
  F.~Domingo, U.~Ellwanger and M.~A.~Sanchis-Lozano,
  Phys.\ Rev.\ Lett.\  {\bf 103} (2009) 111802
  [arXiv:0907.0348 [hep-ph]].
  
\bibitem{Dermisek:2010mg}
  R.~Dermisek and J.~F.~Gunion,
  Phys.\ Rev.\ D {\bf 81} (2010) 075003
  [arXiv:1002.1971 [hep-ph]].
  
\bibitem{Domingo:2010am}
  F.~Domingo,
  JHEP {\bf 1104} (2011) 016
  [arXiv:1010.4701 [hep-ph]].
  
\bibitem{Almarashi:2012ri}
  M.~M.~Almarashi and S.~Moretti,
  arXiv:1205.1683 [hep-ph].
  
\bibitem{Rathsman:2012dp}
  J.~Rathsman and T.~Rossler,
  Adv.\ High Energy Phys.\  {\bf 2012} (2012) 853706
  [arXiv:1206.1470 [hep-ph]].
  
\bibitem{Cerdeno:2013cz}
  D.~G.~Cerdeno, P.~Ghosh and C.~B.~Park,
  JHEP {\bf 1306} (2013) 031
  [arXiv:1301.1325 [hep-ph]].
  
\bibitem{Bomark:2014gya}
  N.~E.~Bomark, S.~Moretti, S.~Munir and L.~Roszkowski,
  JHEP {\bf 1502} (2015) 044
  [arXiv:1409.8393 [hep-ph]].
  
\bibitem{Bomark:2015fga}
  N.~E.~Bomark, S.~Moretti and L.~Roszkowski,
  J.\ Phys.\ G {\bf 43} (2016) no.10,  105003
  [arXiv:1503.04228 [hep-ph]].
  
\bibitem{Conte:2016zjp}
  E.~Conte, B.~Fuks, J.~Guo, J.~Li and A.~G.~Williams,
  JHEP {\bf 1605} (2016) 100
  [arXiv:1604.05394 [hep-ph]].
  
\bibitem{Aad:2012tfa}
  G.~Aad {\it et al.} [ATLAS Collaboration],
  Phys.\ Lett.\ B {\bf 716} (2012) 1
  [arXiv:1207.7214 [hep-ex]].
  
\bibitem{Chatrchyan:2012xdj}
  S.~Chatrchyan {\it et al.} [CMS Collaboration],
  Phys.\ Lett.\ B {\bf 716} (2012) 30
  [arXiv:1207.7235 [hep-ex]].
  
\bibitem{Cao:2013gba}
  J.~Cao, F.~Ding, C.~Han, J.~M.~Yang and J.~Zhu,
  JHEP {\bf 1311} (2013) 018
  [arXiv:1309.4939 [hep-ph]].
  
\bibitem{Aad:2015oqa}
  G.~Aad {\it et al.} [ATLAS Collaboration],
  Phys.\ Rev.\ D {\bf 92} (2015) no.5,  052002
  [arXiv:1505.01609 [hep-ex]].
  
\bibitem{Khachatryan:2015wka}
  V.~Khachatryan {\it et al.} [CMS Collaboration],
  Phys.\ Lett.\ B {\bf 752} (2016) 146
  [arXiv:1506.00424 [hep-ex]].
  
\bibitem{Khachatryan:2015nba}
  V.~Khachatryan {\it et al.} [CMS Collaboration],
  JHEP {\bf 1601} (2016) 079
  [arXiv:1510.06534 [hep-ex]].
  
\bibitem{Cerdeno:2004xw}
  D.~G.~Cerdeno, C.~Hugonie, D.~E.~Lopez-Fogliani, C.~Munoz and A.~M.~Teixeira,
  JHEP {\bf 0412} (2004) 048
  [hep-ph/0408102].
  
\bibitem{Belanger:2005kh}
  G.~Belanger, F.~Boudjema, C.~Hugonie, A.~Pukhov and A.~Semenov,
  JCAP {\bf 0509} (2005) 001
  [hep-ph/0505142].
  
\bibitem{Cerdeno:2007sn}
  D.~G.~Cerdeno, E.~Gabrielli, D.~E.~Lopez-Fogliani, C.~Munoz and A.~M.~Teixeira,
  JCAP {\bf 0706} (2007) 008
  [hep-ph/0701271 [HEP-PH]].
  
\bibitem{Hugonie:2007vd}
  C.~Hugonie, G.~Belanger and A.~Pukhov,
  JCAP {\bf 0711} (2007) 009
  [arXiv:0707.0628 [hep-ph]].
  
\bibitem{Vasquez:2010ru}
  D.~Albornoz Vasquez, G.~Belanger, C.~Boehm, A.~Pukhov and J.~Silk,
  Phys.\ Rev.\ D {\bf 82} (2010) 115027
  [arXiv:1009.4380 [hep-ph]].
  
\bibitem{Cao:2011re}
  J.~J.~Cao, K.~i.~Hikasa, W.~Wang, J.~M.~Yang, K.~i.~Hikasa, W.~Y.~Wang and J.~M.~Yang,
  Phys.\ Lett.\ B {\bf 703} (2011) 292
  [arXiv:1104.1754 [hep-ph]].
  
\bibitem{Vasquez:2012hn}
  D.~Albornoz Vasquez, G.~Belanger, C.~Boehm, J.~Da Silva, P.~Richardson and C.~Wymant,
  Phys.\ Rev.\ D {\bf 86} (2012) 035023
  [arXiv:1203.3446 [hep-ph]].

\bibitem{Han:2014nba}
  T.~Han, Z.~Liu and S.~Su,
  JHEP {\bf 1408} (2014) 093
  doi:10.1007/JHEP08(2014)093
  [arXiv:1406.1181 [hep-ph]].
  
\bibitem{Ellwanger:2014dfa}
  U.~Ellwanger and C.~Hugonie,
  JHEP {\bf 1408} (2014) 046
  [arXiv:1405.6647 [hep-ph]].
  
\bibitem{Han:2015zba}
  C.~Han, D.~Kim, S.~Munir and M.~Park,
  JHEP {\bf 1507} (2015) 002
  [arXiv:1504.05085 [hep-ph]].
  
\bibitem{Cerdeno:2015jca}
  D.~G.~Cerdeno, M.~Peiro and S.~Robles,
  JCAP {\bf 1604} (2016) no.04,  011
  [arXiv:1507.08974 [hep-ph]].
  
\bibitem{Ellwanger:2016wfe}
  U.~Ellwanger and S.~Moretti,
  JHEP {\bf 1611} (2016) 039
  [arXiv:1609.01669 [hep-ph]].
  
\bibitem{Krasznahorkay:2015iga}
  A.~J.~Krasznahorkay {\it et al.},
  Phys.\ Rev.\ Lett.\  {\bf 116} (2016) no.4,  042501
  [arXiv:1504.01527 [nucl-ex]].
  
\bibitem{Dolan:2014ska}
  M.~J.~Dolan, F.~Kahlhoefer, C.~McCabe and K.~Schmidt-Hoberg,
  JHEP {\bf 1503} (2015) 171
   Erratum: [JHEP {\bf 1507} (2015) 103]
  [arXiv:1412.5174 [hep-ph]].
  
\bibitem{Domingo:2011rn}
  F.~Domingo and U.~Ellwanger,
  JHEP {\bf 1106} (2011) 067
  [arXiv:1105.1722 [hep-ph]].
  
\bibitem{Ellwanger:2016qax}
  U.~Ellwanger and C.~Hugonie,
  JHEP {\bf 1605} (2016) 114
  [arXiv:1602.03344 [hep-ph]].
  
\bibitem{Domingo:2016unq}
  F.~Domingo, S.~Heinemeyer, J.~S.~Kim and K.~Rolbiecki,
  Eur.\ Phys.\ J.\ C {\bf 76} (2016) no.5,  249
  [arXiv:1602.07691 [hep-ph]].
  
\bibitem{Rosenzweig:1979ay}
  C.~Rosenzweig, J.~Schechter and C.~G.~Trahern,
  Phys.\ Rev.\ D {\bf 21} (1980) 3388.
  
 \bibitem{DiVecchia:1980yfw}
  P.~Di Vecchia and G.~Veneziano,
  Nucl.\ Phys.\ B {\bf 171} (1980) 253.
  
\bibitem{Kawarabayashi:1980dp}
  K.~Kawarabayashi and N.~Ohta,
  Nucl.\ Phys.\ B {\bf 175} (1980) 477.
  
\bibitem{Kawarabayashi:1980uh}
  K.~Kawarabayashi and N.~Ohta,
  Prog.\ Theor.\ Phys.\  {\bf 66} (1981) 1789.
  
\bibitem{Gasser:1984gg}
  J.~Gasser and H.~Leutwyler,
  Nucl.\ Phys.\ B {\bf 250} (1985) 465.
  
\bibitem{Ellis:1975ap}
  J.~R.~Ellis, M.~K.~Gaillard and D.~V.~Nanopoulos,
  Nucl.\ Phys.\ B {\bf 106} (1976) 292.
  
\bibitem{Shifman:1979eb}
  M.~A.~Shifman, A.~I.~Vainshtein, M.~B.~Voloshin and V.~I.~Zakharov,
  Sov.\ J.\ Nucl.\ Phys.\  {\bf 30} (1979) 711
   [Yad.\ Fiz.\  {\bf 30} (1979) 1368].
  
\bibitem{Vainshtein:1980ea}
  A.~I.~Vainshtein, V.~I.~Zakharov and M.~A.~Shifman,
  Sov.\ Phys.\ Usp.\  {\bf 23} (1980) 429
   [Usp.\ Fiz.\ Nauk {\bf 131} (1980) 537].

\bibitem{Voloshin:1985tc}
  M.~B.~Voloshin,
  Sov.\ J.\ Nucl.\ Phys.\  {\bf 44} (1986) 478
   [Yad.\ Fiz.\  {\bf 44} (1986) 738].
  
\bibitem{Voloshin:1986hp}
  M.~B.~Voloshin,
  Sov.\ J.\ Nucl.\ Phys.\  {\bf 45} (1987) 122
   [Yad.\ Fiz.\  {\bf 45} (1987) 190].
  
\bibitem{Ruskov:1987jg}
  R.~Ruskov,
  Phys.\ Lett.\ B {\bf 187} (1987) 165.
  
\bibitem{Grinstein:1988yu}
  B.~Grinstein, L.~J.~Hall and L.~Randall,
  Phys.\ Lett.\ B {\bf 211} (1988) 363.
  
\bibitem{Chivukula:1988gp}
  R.~S.~Chivukula and A.~V.~Manohar,
  Phys.\ Lett.\ B {\bf 207} (1988) 86
   Erratum: [Phys.\ Lett.\ B {\bf 217} (1989) 568].
  
\bibitem{Chivukula:1989ds}
  R.~S.~Chivukula, A.~G.~Cohen, H.~Georgi, B.~Grinstein and A.~V.~Manohar,
  Annals Phys.\  {\bf 192} (1989) 93.
  
\bibitem{Leutwyler:1989tn}
  H.~Leutwyler and M.~A.~Shifman,
  Phys.\ Lett.\ B {\bf 221} (1989) 384.
  
\bibitem{Dawson:1989kr}
  S.~Dawson,
  Phys.\ Lett.\ B {\bf 222} (1989) 143.
  
\bibitem{Leutwyler:1989xj}
  H.~Leutwyler and M.~A.~Shifman,
  Nucl.\ Phys.\ B {\bf 343} (1990) 369.
  
\bibitem{Pich:1991dg}
  A.~Pich, J.~Prades and P.~Yepes,
  Nucl.\ Phys.\ B {\bf 388} (1992) 31.
  
\bibitem{Grzadkowski:1992av}
  B.~Grzadkowski and J.~Pawelczyk,
  Phys.\ Lett.\ B {\bf 300} (1993) 387.
  
\bibitem{Chang:2008np}
  Q.~Chang and Y.~D.~Yang,
  Phys.\ Lett.\ B {\bf 676} (2009) 88
  [arXiv:0808.2933 [hep-ph]].
  
\bibitem{diCortona:2015ldu}
  G.~Grilli di Cortona, E.~Hardy, J.~Pardo Vega and G.~Villadoro,
  JHEP {\bf 1601} (2016) 034
  [arXiv:1511.02867 [hep-ph]].
  
\bibitem{Wess:1971yu}
  J.~Wess and B.~Zumino,
  Phys.\ Lett.\  {\bf 37B} (1971) 95.
  doi:10.1016/0370-2693(71)90582-X
  
\bibitem{Witten:1983tw}
  E.~Witten,
  Nucl.\ Phys.\ B {\bf 223} (1983) 422.
  doi:10.1016/0550-3213(83)90063-9
  
\bibitem{Ambrosino:2009sc}
  F.~Ambrosino {\it et al.},
  JHEP {\bf 0907} (2009) 105
  [arXiv:0906.3819 [hep-ph]].
  
\bibitem{Ricciardi:2012xu}
  G.~Ricciardi,
  Phys.\ Rev.\ D {\bf 86} (2012) 117505
  [arXiv:1209.3386 [hep-ph]].
  
\bibitem{Pham:2015ina}
  T.~N.~Pham,
  Phys.\ Rev.\ D {\bf 92} (2015) no.5,  054021
  [arXiv:1504.05414 [hep-ph]].
  
\bibitem{Olive:2016xmw}
  C.~Patrignani {\it et al.} [Particle Data Group Collaboration],
  Chin.\ Phys.\ C {\bf 40} (2016) no.10,  100001.
  
\bibitem{Gerard:2005yk}
  J.~M.~Gerard, C.~Smith and S.~Trine,
  Nucl.\ Phys.\ B {\bf 730} (2005) 1
  [hep-ph/0508189].
  
\bibitem{Abouzaid:2006kk}
  E.~Abouzaid {\it et al.} [KTeV Collaboration],
  Phys.\ Rev.\ D {\bf 75} (2007) 012004
  [hep-ex/0610072].
  
\bibitem{Dorokhov:2007bd}
  A.~E.~Dorokhov and M.~A.~Ivanov,
  Phys.\ Rev.\ D {\bf 75} (2007) 114007
  [arXiv:0704.3498 [hep-ph]].
  
\bibitem{Cronin:1967jq}
  J.~A.~Cronin,
  Phys.\ Rev.\  {\bf 161} (1967) 1483.
 
\bibitem{Osborn:1970nn}
  H.~Osborn and D.~J.~Wallace,
  Nucl.\ Phys.\ B {\bf 20} (1970) 23.

\bibitem{Holstein:2001bt}
  B.~R.~Holstein,
  Phys.\ Scripta T {\bf 99} (2002) 55
  [hep-ph/0112150].
  
\bibitem{Gunion:1989we}
  J.~F.~Gunion, H.~E.~Haber, G.~L.~Kane and S.~Dawson,
  Front.\ Phys.\  {\bf 80} (2000) 1.
  
\bibitem{NMSSMTools}
  U.~Ellwanger, J.~F.~Gunion and C.~Hugonie,
  JHEP {\bf 0502} (2005) 066
  [hep-ph/0406215];
\\
  U.~Ellwanger and C.~Hugonie,
  Comput.\ Phys.\ Commun.\  {\bf 175} (2006) 290
  [hep-ph/0508022];
\\
  U.~Ellwanger, C.~Hugonie, {\it Comput.~Phys.~Commun.}
  {\bf 175} (2006) 290, arXiv:hep-ph/0508022;\\ 
\verb|http://www.th.u-psud.fr/NMHDECAY/nmssmtools.html|~.
 
\end{thebibliography}
\end{document}